\documentclass[onecolumn,pre,tightenlines,superscriptaddress,floatfix]{revtex4-2}
\usepackage{hyperref}
\usepackage{amsmath,amssymb}
\usepackage[utf8]{inputenc}
\usepackage[T1]{fontenc}
\usepackage{graphicx}
\graphicspath{{generated/figures/}}
\usepackage{float}
\usepackage[usenames,dvipsnames]{xcolor}
\usepackage[normalem]{ulem}
\definecolor{RED}{named}{red}
\usepackage{tikz}
\usepackage{mathtools}
\usepackage{booktabs}
\usepackage{siunitx}
\usepackage{multirow}
\usepackage{subcaption}
\usepackage{verbatim}
\usepackage{stmaryrd}
\usepackage{tikz}
\newcommand*\circled[1]{\tikz[baseline=(char.base)]{
            \node[shape=circle,draw,inner sep=0.8pt] (char) {#1};}}

\definecolor{warningcolor}{rgb}{1,0.5,0}

\newcommand{\warning}[1]{}

\begin{document}

\title{\texorpdfstring{Hierarchical Bayesian calibration of mesoscopic models for ultrasound contrast agents from force spectroscopy data}{Hierarchical Bayesian calibration of ultrasound contrast agents mesoscopic models from force spectroscopy data}}

\author{Brieuc Benvegnen}
\affiliation{Universitat de Barcelona Institute of Complex Systems (UBICS), C/ Martí i Franqués 1, 08028 Barcelona, Spain}

\author{Nikolaos Ntarakas}
\affiliation{Theory Department, National Institute of Chemistry, Hajdrihova 19, SI-1001 Ljubljana, Slovenia}
\affiliation{Department of Physics, Faculty of Mathematics and Physics, University of Ljubljana, Jadranska 19, SI-1000 Ljubljana, Slovenia}

\author{Tilen Potisk}
\affiliation{Theory Department, National Institute of Chemistry, Hajdrihova 19, SI-1001 Ljubljana, Slovenia}
\affiliation{Department of Physics, Faculty of Mathematics and Physics, University of Ljubljana, Jadranska 19, SI-1000 Ljubljana, Slovenia}

\author{Ignacio Pagonabarraga}
\affiliation{Universitat de Barcelona Institute of Complex Systems (UBICS), C/ Martí i Franqués 1, 08028 Barcelona, Spain}

\author{Matej Praprotnik}
\email{praprot@cmm.ki.si}
\thanks{Corresponding author.}
\affiliation{Universitat de Barcelona Institute of Complex Systems (UBICS), C/ Martí i Franqués 1, 08028 Barcelona, Spain}
\affiliation{Theory Department, National Institute of Chemistry, Hajdrihova 19, SI-1001 Ljubljana, Slovenia}
\affiliation{Department of Physics, Faculty of Mathematics and Physics, University of Ljubljana, Jadranska 19, SI-1000 Ljubljana, Slovenia}

\date{\today}

\begin{abstract}
Ultrasound-guided drug and gene delivery (\textsc{usdg}) is a promising non-invasive bioengineering approach for targeted therapeutic applications. Mechanical properties of encapsulated microbubbles (\textsc{emb}s), which serve as ultrasound contrast agents and targeted drug delivery vehicles, strongly affect their specific interactions with ultrasound and are thus critical to the success and efficiency of \textsc{usdg}. Accurate calibration of high-fidelity numerical particle-based models of \textsc{emb} capsid mechanics, with quantified uncertainty, is computationally challenging because direct Bayesian inference with dissipative particle dynamics (\textsc{dpd}) is prohibitively expensive. We employ a surrogate-accelerated Bayesian calibration workflow that combines deep neural network (\textsc{dnn}) surrogates, transitional Markov chain Monte Carlo sampling, and hierarchical regularization across \textsc{emb} diameters. The surrogate models reproduce the \textsc{dpd} responses with low validation error and provide large computational speedups relative to direct simulation, rendering posterior sampling tractable. The workflow first uses \textsc{dpd} simulations to train and validate the \textsc{dnn} surrogates, then uses the full \textsc{emb} model for sensitivity analysis, carries a reduced \textsc{emb} model, in which weakly contributing nonlinear elastic energy terms are omitted, through independent and hierarchical inference, and finally re-evaluates the full model, with displacement offset and measurement noise treated as separate calibration parameters. Using this framework, we develop two data-informed \textsc{dpd} models of commercial \textsc{emb} agents, i.e., Definity and SonoVue, and perform inference of force field parameters based on published compression experiments for Definity and indentation experiments for SonoVue, each spanning three distinct diameters.  The inferred posteriors show that key model parameters, such as the stretching stiffness and bending modulus, are consistently constrained by the available data, while reduced models show close agreement with the full \textsc{emb} model in predictive accuracy. 
The presented methodology can be used to derive bespoke, data-informed models for a wide range of ultrasound contrast agents, including encapsulated gas vesicles, \textsc{emb}s  with diverse capsids consisting of lipids, proteins, or polymers, and functionalized with ligands. Our results show that the surrogate-accelerated Hierarchical Bayesian workflow can calibrate high-fidelity numerical \textsc{emb} models from quasi-static force spectroscopy data with quantified uncertainty, and provide a pathway toward accurate force fields for large-scale simulations of \textsc{usdg}.

\end{abstract}

\maketitle

\setcounter{section}{0}
\makeatletter
\renewcommand{\thesection}{\arabic{section}}
\renewcommand{\thesubsection}{\thesection.\arabic{subsection}}
\renewcommand{\thesubsubsection}{\thesubsection.\arabic{subsubsection}}
\renewcommand{\p@subsection}{}
\renewcommand{\p@subsubsection}{}
\makeatother

\setcounter{figure}{0}
\renewcommand{\thefigure}{\arabic{figure}}

\setcounter{table}{0}
\renewcommand{\thetable}{\arabic{table}}

\section{Introduction}

Encapsulated microbubbles (\textsc{emb}s) are a class of engineered soft materials with gas cores surrounded by elastic shells, composed of a lipid, protein or polymer shell and typically 1 to 10\,\textmu m in diameter~\citep{stride2009physical,ferrara2007ultrasound}. These structures serve dual roles in biomedical applications: as ultrasound contrast agents that enhance diagnostic imaging through acoustic scattering~\citep{klibanov2006microbubble,datta2008ultrasound}, and as targeted drug delivery vehicles capable of payload release upon ultrasonic excitation~\citep{ferrara2007ultrasound,sirsi2014state,stride2010cavitation,coussios2008applications}. The mechanical properties of \textsc{emb} shells, particularly their elastic modulus, bending rigidity, and response to compression, critically determine bubble stability under physiological flow conditions, acoustic resonance behavior, and rupture thresholds for controlled drug release~\citep{stride2009physical,guo2013investigation,stride2010cavitation,konofagou2012optimization}. Accurate characterization of these mechanical properties enables rational design of \textsc{emb}s optimized for specific clinical applications, including targeted chemotherapy delivery to tumor vasculature~\citep{tinkov2009microbubbles,tinkov2010new} and blood--brain barrier permeabilization for neurotherapeutics~\citep{hynynen2001noninvasive}. However, systematic quantification of \textsc{emb} mechanical parameters with rigorous uncertainty bounds remains an open challenge, limiting predictive modeling of \textsc{emb} behavior in complex in vivo environments.

\begin{figure}
    \centering
    \includegraphics[width=1.0\linewidth]{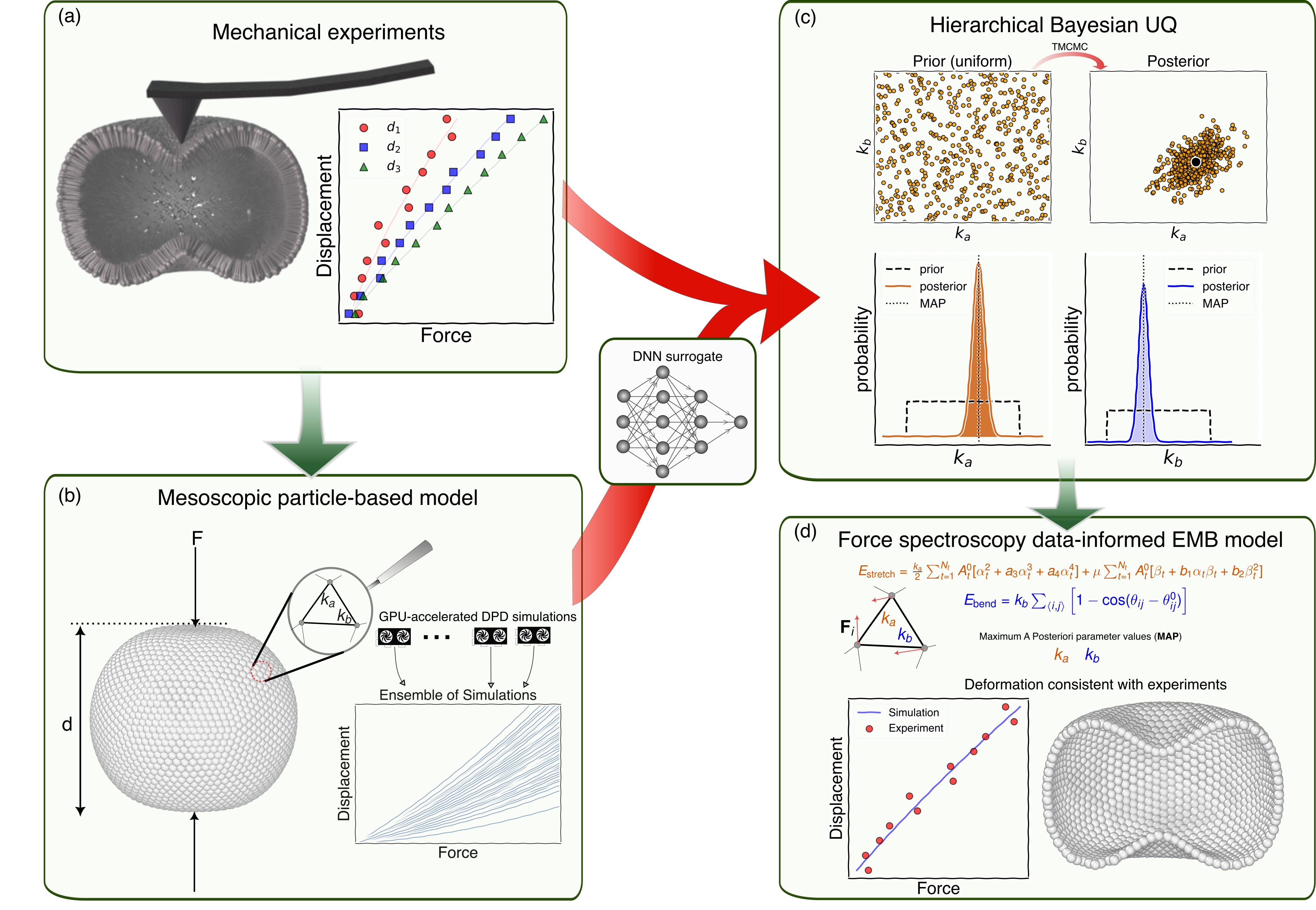}
    \caption{\textbf{Hierarchical Bayesian \textsc{uq}-driven multiscale workflow from experiments through large-scale simulations and deep neural network (\textsc{dnn}) surrogates to a data-informed \textsc{emb} model.} (a) Mechanical experiments, such as indentation or compression, capture the response of soft microstructures and provide datasets in the form of force-displacement curves. (b) A mesoscopic model \cite{ntarakas2025dissipative}, solved using \textsc{gpu}-accelerated simulations, reproduces this behavior through effective model parameters, generating an ensemble of responses across the parameter space. (c) A hierarchical Bayesian uncertainty quantification (\textsc{uq}) workflow, combined with \textsc{dnn} surrogates (based on the ensemble of generated simulation data), enables efficient parameter inference from experimental data with quantified uncertainty. (d) The inferred posteriors, specifically the Maximum A Posteriori (\textsc{map}) values of the probability distributions, are then used to build the optimal data-informed \textsc{emb} model, which can be used in realistic large-scale simulations.}
    \label{fig:overview}
\end{figure}

Experimental characterization of \textsc{emb} mechanical properties typically employs atomic force microscopy (\textsc{afm}) compression~\citep{glynos2009polymeric,glynos2009nanomechanics}, micropipette aspiration~\citep{evans1989aspiration} or ultrasonic attenuation measurements~\citep{church1995effects,coussios2008applications}. \textsc{afm} compression provides direct force-displacement relationships at nanometer resolution but is subject to experimental uncertainties, including contact point determination, tip-sample alignment artifacts, and batch-to-batch variability in \textsc{emb} synthesis~\cite{glynos2009nanomechanics,tu2009estimating}. Translating these force-displacement curves into fundamental material properties (e.g., Young's modulus, bending modulus) requires computational models that can simulate \textsc{emb} deformation under realistic experimental conditions. Continuum models based on finite element methods offer computational efficiency and have been widely used to model \textsc{emb} dynamics in fluid and vascular environments~\citep{zhang2018mechanics,guo2024ultrasound}, but make strong assumptions about shell constitutive behavior (e.g., Neo-Hookean elasticity) and cannot capture heterogeneities at the molecular scale of shell composition. Particle-based mesoscopic simulations, particularly dissipative particle dynamics (\textsc{dpd}), provide a complementary approach that explicitly represents the shell microstructure through coarse-grained particles interacting via prescribed force fields~\citep{groot1997dissipative,espanol1995statistical}. Our previous work demonstrated that \textsc{dpd} models with isotropic elasticity and bending resistance can accurately reproduce \textsc{emb} compression behavior across multiple shell-chemistry compositions~\citep{ntarakas2025dissipative}. However, \textsc{dpd} simulations require 30--60 minutes per parameter configuration on modern \textsc{gpu}s (e.g. \textsc{nvidia} A40), making systematic parameter space exploration prohibitively expensive. A single Bayesian inference campaign that requires $10^5$ likelihood evaluations would already consume roughly $3\times10^{6}$-- $6\times10^{6}$~\textsc{gpu}-minutes, i.e. about 5.7--11.4 \textsc{gpu}-years, even before accounting for repeated phases, restarts, or the one-time generation of surrogate-training data, making rigorous uncertainty quantification (\textsc{uq}) intractable without methodological innovations. 
\textsc{uq} constitutes a framework for addressing imperfections in models, simulations, and experimental data \citep{angelikopoulos2012bayesian}. By representing unknown parameters and model discrepancies as distributions instead of fixed values, \textsc{uq} enables predictions that reflect both expected behavior and associated confidence, which is essential for making reliable, data-informed decisions in complex systems. Bayesian inference serves as a principled framework for \textsc{uq} through parameter estimation with quantified uncertainties~\citep{gelman2013bayesian,gregory2005bayesian,nordman2025bayesian,amoudruz2025bayesian}, transforming point estimates into probability distributions that capture both data informativeness and prior knowledge. Hierarchical Bayesian \textsc{uq} extends this framework to settings with multiple datasets or experimental conditions, allowing parameters to vary across conditions while sharing a common structure \citep{wu2016fusing}. This enables principled information sharing, improves parameter identifiability, and yields a more realistic representation of variability in complex systems. In molecular and mesoscopic simulations, Bayesian approaches have been used to calibrate force fields for proteins~\citep{cailliez2014calibration}, polymers~\citep{zhang2014polymer}, and soft materials~\citep{angelikopoulos2012bayesian}. The computational bottleneck of repeated forward model evaluations can be mitigated with surrogate modeling, using fast approximations such as Gaussian processes, neural networks, or polynomial chaos expansions that learn input-output relationships from a limited training dataset~\citep{rasmussen2006gaussian,wu2022multi}. Deep neural network (\textsc{dnn}) surrogates are particularly attractive for high-dimensional nonlinear systems because they can reproduce complex input-output maps with large speedups over direct simulation~\citep{raissi2019physics,lu2021deepxde}. Bayesian optimization has also emerged as an efficient strategy for refining model parameters in computationally expensive settings, where each evaluation requires a forward simulation \citep{ray2025refining}. By combining a surrogate model with an acquisition function, it enables adaptive exploration of the parameter space and rapid identification of optimal coarse-grained representations with minimal simulation cost. For Bayesian sampling, the transitional Markov chain Monte Carlo (\textsc{tmcmc}) methods~\citep{ching2007transitional,beck2002bayesian} efficiently explore complex posterior distributions through a sequence of intermediate distributions. When multiple experimental datasets are available, hierarchical Bayesian models enable principled information sharing across conditions while preserving dataset-specific variability~\citep{gelman2013bayesian,gelman2006multilevel,economides2021hierarchical}. Together, these components establish a practical route for uncertainty-aware calibration of particle-based \textsc{emb} models. Previous Bayesian approaches in molecular simulations assess predictive performance by verifying whether experimental observations fall within posterior predictive confidence intervals \cite{economides2021hierarchical}. Although this offers a qualitative measure of agreement, it does not quantify the degree of calibration. Here, we use coverage only as one descriptive posterior-predictive agreement diagnostic under the current heteroscedastic Gaussian likelihood, defined as the fraction of experimental observations contained within the pointwise $90\%$ posterior predictive interval. In Results, this interval-inclusion diagnostic is complemented by a proper scoring-rule comparison of the full pointwise posterior predictive distributions. \newline
The objective of this study is to calibrate a high-fidelity numerical \textsc{emb} model from compression and indentation data with quantified uncertainty, (See Fig.~\ref{fig:overview} for a schematic overview of the entire workflow). To make this feasible, we combine \textsc{dnn} surrogates trained on \textsc{dpd} simulations with \textsc{tmcmc} sampling, and use hierarchical modeling to regularize inference across different \textsc{emb} diameters. The framework first uses surrogate-based sensitivity analysis to determine which membrane parameters are actually informed by the present quasi-static data. On that basis, the inference workflow proceeds with a reduced membrane model in which $(k_a,k_b)$ are the shell-mechanics parameters, while $d_0$ and measurement noise are treated separately as calibration and statistical quantities. Independent single-diameter inference is then compared with hierarchical inference that shares information across diameters, and the full nonlinear force field is rerun only at the end as a confirmatory safety check. \newline
We first validate the surrogates and then use the full model only to identify which parameters are relevant in the present quasi-static regime. Next, we infer reduced-model posteriors independently for each diameter, pool information hierarchically across diameters, propagate the resulting uncertainty to the response curves, compare representative direct \textsc{dpd} checks with the experimental data, and finally rerun the full nonlinear model as a last sanity check. The Methods section describes the \textsc{dpd} simulations, experimental-data preprocessing, surrogate construction, hierarchical Bayesian formulation, uncertainty propagation, and computational implementation used throughout the study. Our central claim is that a surrogate-accelerated Bayesian workflow can calibrate a high-fidelity numerical \textsc{emb} model with quantified uncertainty and that, in the present quasi-static regime, the data constrain $k_a$ and $k_b$ and support a reduced force field description. That conclusion rests on one explicit chain of evidence: dominant $k_a/k_b$ sensitivities, stable independent and hierarchical reduced posteriors, close full-vs-reduced \textsc{map} overlays, and close reduced-vs-full posterior-predictive continuous ranked probability score (\textsc{crps}).

\section{Results}
\label{sec:results}

Results are reported for two \textsc{emb} models of commercial agents with lipid shells (Definity and SonoVue) and three bubble diameters per \textsc{emb} model. Definity models are calibrated using experimental data from \textsc{afm} compression, whereas SonoVue models are calibrated with data from \textsc{afm} indentation. The inference conditions on one reprocessed published curve per diameter; the exact source references and retained point counts are summarized in Table~\ref{tab:data_summary} in Supplementary Information. We first use the full membrane force field only to establish the sensitivity structure of the inverse problem and then, from the sensitivity section onward, carry a reduced model with membrane parameters $(k_a,k_b)$ through the main-text inference workflow, treating $d_0$ and $\sigma$ separately as calibration and noise parameters. The full model is rerun only in the final sanity-check section. 

\subsection{\texorpdfstring{\textsc{dpd}}{DPD} force field and \textsc{dnn} surrogate validation}
\label{sec:res:dpd_surrogate}

\begin{figure}[t]
\centering
\includegraphics[width=0.98\linewidth]{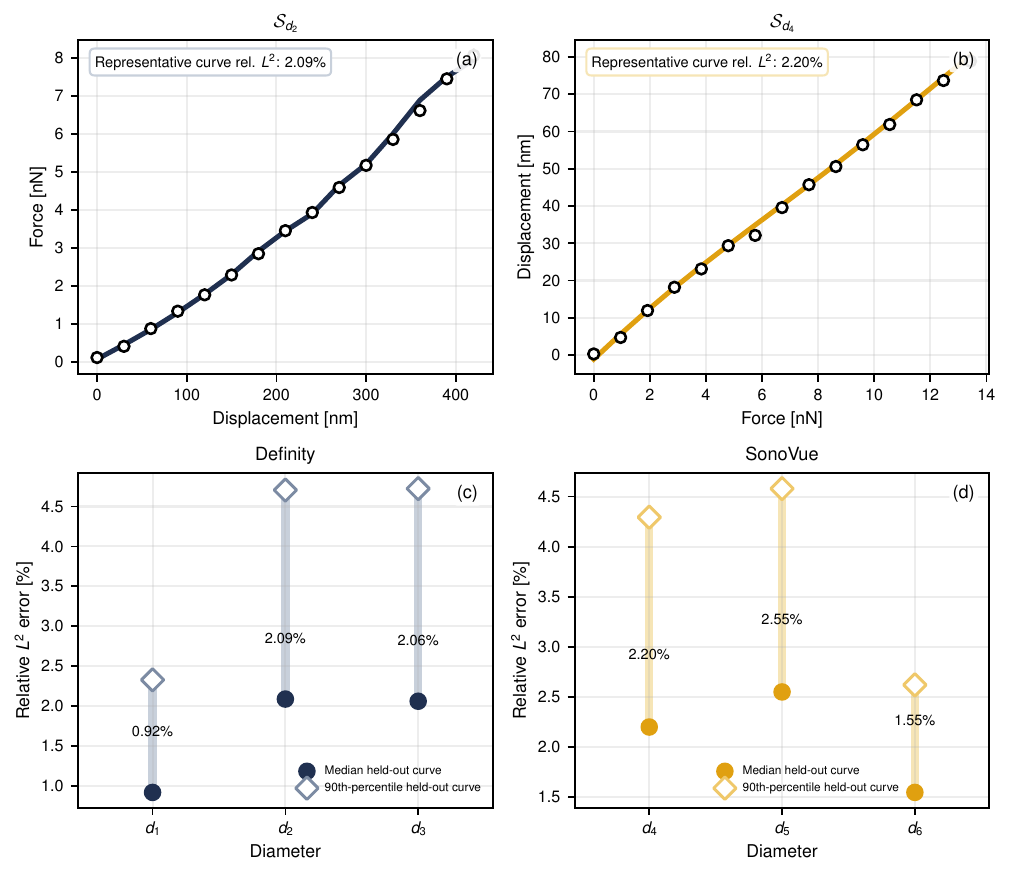}
\caption{\textbf{Grouped-holdout surrogate validation.} Panels (a–b) compare held-out \textsc{dpd} curves with the corresponding \textsc{dnn} predictions for representative $\mathcal{S}_{d_2}$ (ink) and $\mathcal{S}_{d_4}$ (gold) surrogates. Panels (c–d) summarize, for each \textsc{emb} model, the median and 90th-percentile held-out-curve relative $L^2$ errors from the grouped parameter-set holdout audit.}
\label{fig:surrogate_validation_curves}
\end{figure}

We consider six \textsc{dpd} models $(\mathcal{M}_{d_i})_{i \in \llbracket 1,6 \rrbracket}$, which share the same physical formulation and parameterization~\citep{ntarakas2025dissipative}, but correspond to different experimental conditions (See Fig.~\ref{fig:workflow}, block \circled{1}). Models $\mathcal{M}_{d_1}, \mathcal{M}_{d_2}, \mathcal{M}_{d_3}$ describe Definity bubbles~\cite{lindner2002microvascular,datta2008ultrasound,baseri2010multi} (compression experiments) for fixed diameters ($d_1, d_2, d_3$) respectively, while $\mathcal{M}_{d_4}, \mathcal{M}_{d_5}, \mathcal{M}_{d_6}$ describe SonoVue bubbles~\cite{schneider1999sonovue,demene2021transcranial,tu2009estimating,guo2013investigation,de2007compression} (indentation experiments) for fixed diameters ($d_4, d_5, d_6$) respectively. Throughout, the index $d_i$ denotes the bubble diameter, with $(d_i)_{i=1,\dots,6} = (2.1,\,2.9,\,3.0,\,3.2,\,3.4,\,5.8)\,\mu\mathrm{m}$. We simulate the \textsc{dpd} models using the high-performance \textsc{mirheo} software package \cite{alexeev2020mirheo}. In the compression experiment with Definity, the \textsc{dpd} simulator returns the reaction force as a function of imposed displacement through an explicit plate-contact geometry. In the indentation experiment with SonoVue, it returns the bubble diameter change as a function of applied force under the prescribed loading protocol. These \textsc{dpd} outputs are therefore directly comparable to the corresponding experimental measurements. The membrane force field of each model $\mathcal{M}_{d_i}$ is parameterized by a stretching stiffness $k_a$, a bending modulus $k_b$, nonlinear elastic coefficients $(b_1, b_2, a_3, a_4)$, and a displacement offset $d_0$ used to align the curves. These parameters are collected in the associated vector $\Theta_{d_i} = (k_a, k_b, b_1, b_2, a_3, a_4, d_0)_{d_i}$ while the observation-noise amplitude $\sigma_{d_i}$ is treated separately as a diameter-specific parameter. Full details of the \textsc{dpd} setup and parameter definitions are provided in Methods section.

To enable Bayesian inference at scale, we train \textsc{dnn} surrogates to predict the output of the \textsc{dpd} models on thousands of simulated curves. Each \textsc{dpd} model~$\mathcal{M}_{d_i}$ is subsequently replaced by its surrogate counterpart~$\mathcal{S}_{d_i}$ (See Fig.~\ref{fig:workflow}, block~\circled{2}) during inference, enabling efficient Bayesian calibration. Surrogate selection is performed via an architecture sweep, retaining the \textsc{dnn} with the lowest validation mean-squared error. To assess surrogate accuracy a posteriori, we go further and adopt a stricter grouped parameter-set holdout, excluding entire simulated curves during validation, to complement the pointwise production split~\cite{ma2020eating}. In this unseen-curve audit, the median relative $L^2$ error over the full loading branch remains low: less than 2.1\% for Definity surrogates and less than 2.6\% for SonoVue surrogates (See Supplementary Table~\ref{tab:surrogate_validation}). Fig.~\ref{fig:surrogate_validation_curves} combines representative grouped-holdout curves in real units with a compact across-diameter summary based on the median and 90th-percentile held-out-curve relative $L^2$ errors, making clear that the surrogates remain accurate not only for illustrative held-out curves but also across the bulk of the held-out simulated curves. Representative held-out grouped-validation curves are collected in Supplementary Fig.~\ref{fig:surrogate_group_holdout_examples}. Combined with their orders-of-magnitude speedup over direct simulation, these surrogate errors are low enough to make the subsequent \textsc{tmcmc}-based calibration computationally tractable (Methods).

\subsection{\texorpdfstring{Sensitivity analysis motivates a reduced membrane model}{Sensitivity analysis motivates a reduced membrane model}}
\label{sec:res:sensitivity}

\begin{figure}[t]
\centering
\includegraphics[width=0.98\linewidth]{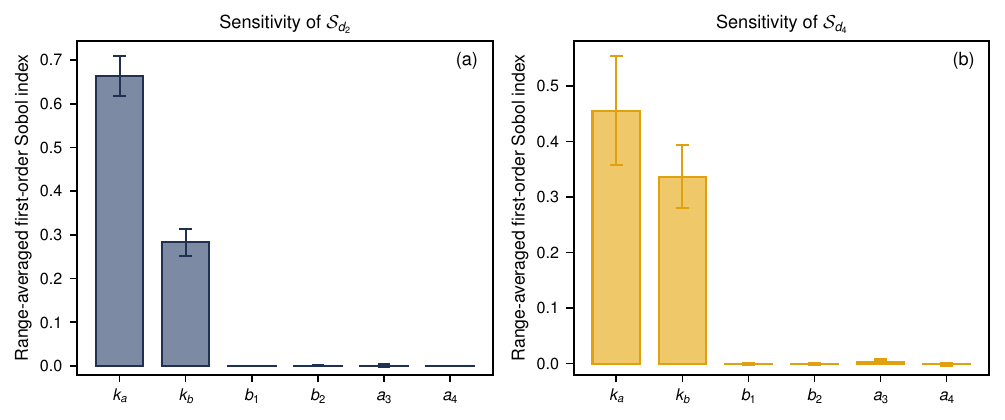}
\caption{\textbf{Range-averaged first-order Sobol sensitivity indices for representative diameters of each \textsc{emb} model}. The bars report the loading-range-averaged first-order index for each material parameter for (a) Definity ($\mathcal{S}_{d_2}$, ink) and (b) SonoVue ($\mathcal{S}_{d_4}$, gold); error bars denote the corresponding averaged Sobol confidence estimates. Only the material parameters are shown.}
\label{fig:sensitivity}
\end{figure}

To determine which force field parameters can be constrained by force spectroscopy data, we quantify the global sensitivities of the surrogates over the experimentally explored range using first-order Sobol indices \cite{herman2017salib}. The resulting hierarchy is clear: the response is governed primarily by the stretching stiffness $k_a$ and the bending modulus $k_b$, while the nonlinear coefficients $(b_1,b_2,a_3,a_4)$ contribute only weakly within the loading regime probed here. Fig.~\ref{fig:sensitivity} reports range-averaged first-order indices for representative surrogates $\mathcal{S}_{d_2}$ and $\mathcal{S}_{d_4}$. For $\mathcal{S}_{d_2}$, the indices assign approximately $66\%$ of the variance to $k_a$ and $28\%$ to $k_b$. For $\mathcal{S}_{d_4}$, the same averaged analysis assigns approximately $45\%$ to $k_a$ and $33\%$ to $k_b$. In both cases, the nonlinear terms remain negligible compared to the two elastic parameters, with bar heights clustered near zero within their confidence intervals. These representative panels are shown because they reflect the same dominant $k_a/k_b$ sensitivity structure observed separately for Definity and SonoVue surrogates.

Collectively, these trends set clear identifiability expectations: in the present deformation regime, the data primarily inform $(k_a,k_b)$, while the nonlinear terms are expected to remain weakly constrained. This motivates the concrete modeling decision adopted in the remainder of this study: after this section, the inference uses a reduced membrane force field with $(k_a,k_b)$ as the shell-mechanics parameters, while $d_0$ and $\sigma$ are retained separately as calibration and noise parameters. In practice, for each bubble $i$, we use the surrogates $\mathcal{S}_{d_i}^0$ obtained by setting the $\mathcal{S}_{d_i}$ nonlinear parameters $(b_1, b_2, a_3, a_4)$ to zero and we use the notation $\theta_{d_i}$ to refer to the associated reduced set of parameters (See Fig.~\ref{fig:workflow}, block \circled{2}): $\theta_{d_i} = (k_a, k_b, d_0)_{d_i}$. We return to the full models only in the final Results subsection as a safety check on this reduction.

\subsection{\texorpdfstring{ Hierarchical Bayesian inference of the reduced model}{Hierarchical Bayesian inference of the reduced model}}
\label{sec:res:phase1}
\begin{figure}[t]
\centering
\includegraphics[width=0.98\textwidth]{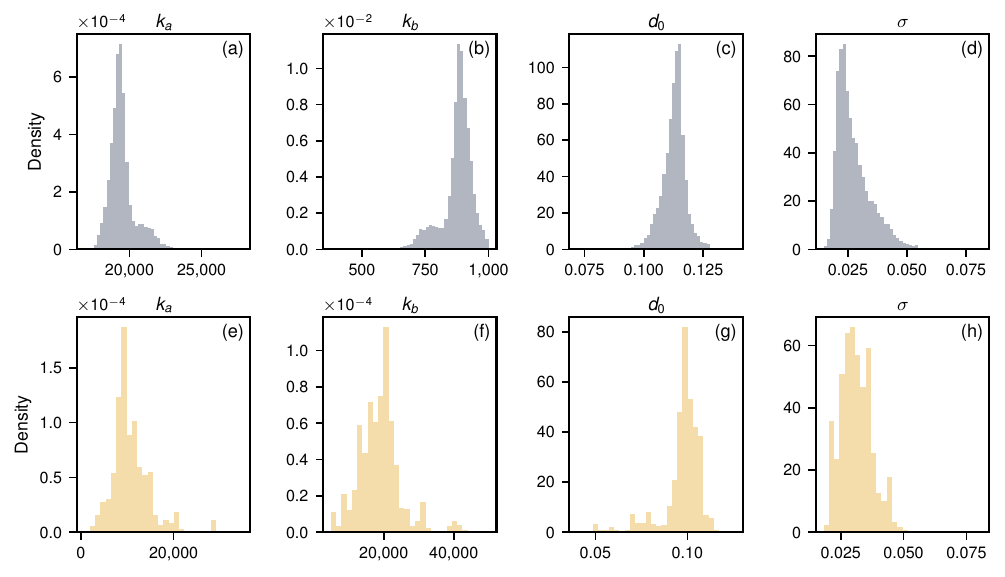}
\caption{\textbf{Representative reduced-model independent posterior marginals for $\mathcal{\theta}_{d_2}$ and $\mathcal{\theta}_{d_4}$.} 
Panels (a–d) correspond to $\mathcal{\theta}_{d_2}$ (ink) and (e–h) to $\mathcal{\theta}_{d_4}$ (gold). 
The reduced independent inferences yield well-defined posteriors for the membrane parameters $(k_a,k_b)$, the calibration parameter $d_0$, and the noise parameter $\sigma$.}
\label{fig:phase1_reduced_representative}
\end{figure}

To establish a results baseline, we first perform independent Bayesian calibrations for each diameter using only the reduced models (See Fig.~\ref{fig:workflow}, block \circled{3}). For each bubble $i$ (compression experiment using Definity: 2.1, 2.9, and 3.0~$\mu$m; indentation experiment using SonoVue: 3.2, 3.4, and 5.8~$\mu$m), we run a separate \textsc{tmcmc} inference using the \textsc{korali} software framework \cite{martin2022korali} and the corresponding \textsc{dnn} surrogate as the forward model, where $d_0$ is an auxiliary curve-alignment/contact-offset parameter and $\sigma$ controls the (relative) observation noise model (Methods).
At this reduced-model level, independent Bayesian calibrations already yield well-defined single-diameter posterior marginals $P(k_a)$, $P(k_b)$, $P(d_0)$, and $P(\sigma)$. The Definity model posteriors are already sharply localized in all four inferred quantities, while the SonoVue model posteriors remain broader but still resolve a stable reduced parameter set. The value of $\sigma$ is on the order of $2.5-5\%$ of the model output, indicating that the inferred observation noise corresponds to a small fraction of the model output across the explored regime. Fig.~\ref{fig:phase1_reduced_representative} depicts representative reduced marginals for $\mathcal{\theta}_{d_2}$ and $\mathcal{\theta}_{d_4}$ and the associated inferred noises. These independent single-diameter calibrations therefore establish the baseline reduced-model information content of each curve before pooling information across diameters.

\label{sec:res:phase2}
\begin{figure}[t]
\centering
\includegraphics[width=\textwidth]{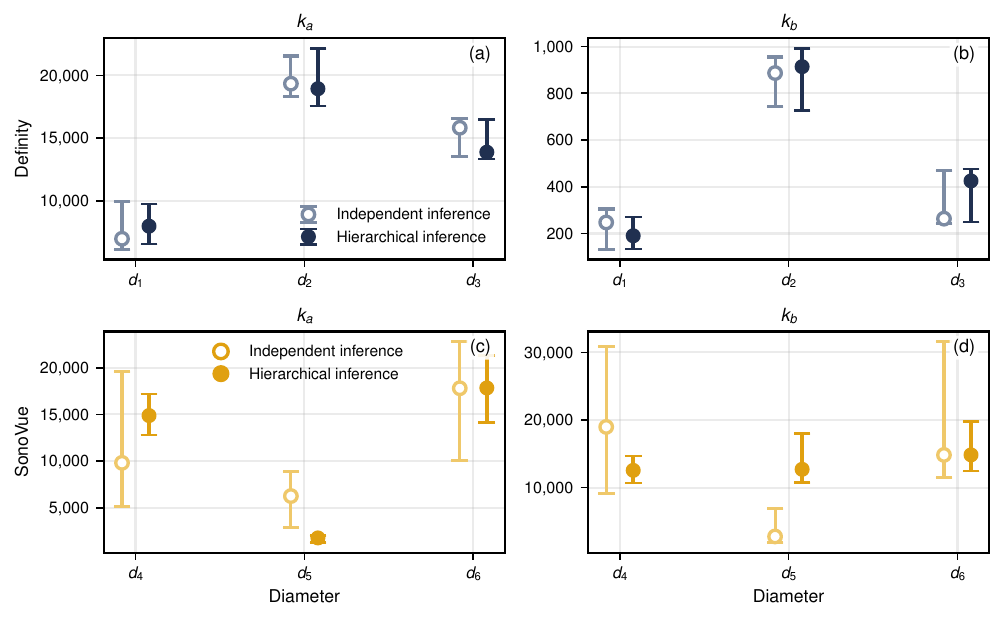}
\caption{\textbf{Reduced-model posterior centers and 90\% intervals for independent and hierarchical inferences for the dominant elastic parameters $k_a$ and $k_b$ (\textsc{dpd} units)}. Each point denotes the posterior median, and each interval denotes the central 90\% range. Panels (a–b) correspond to Definity \textsc{emb}s (ink), and panels (c–d) to SonoVue \textsc{emb}s (gold).
This compact comparison illustrates how the priors learned in the hierarchical inference regularize the final reduced posteriors relative to the independent reduced posteriors.}
\label{fig:phase1_vs_phase3b_reduced_summary}
\end{figure}
For both Definity and SonoVue models, we then infer hierarchical hyperparameters for the reduced models to quantify how $(k_a, k_b, d_0)$ vary across diameters (Methods, Section~\ref{sec:meth:bayes}). For each \textsc{emb} model, this step pools information across its three diameters to learn hierarchical priors that regularize the independently inferred posteriors. Using these learned priors, we subsequently perform diameter-specific \textsc{tmcmc} inference for the three diameters of each \textsc{emb} model (See Fig.~\ref{fig:workflow}, block \circled{3}). For each bubble $i$, these posteriors $P(\tilde\theta_{d_i})$ constitute the final reduced-model distributions used downstream for propagated-response bands, Maximum A Posteriori (\textsc{map}) extraction, and full-model consistency checks. Fig.~\ref{fig:phase1_vs_phase3b_reduced_summary} compares the medians and 90\% credible intervals of these posteriors $P(\tilde\theta_{d_i})$ with those obtained from independent reduced-model inferences $P(\theta_{d_i})$, for the elastic parameters. For Definity models, the hierarchy modifies the posteriors only moderately and preserves the same identifiable region. For SonoVue models, the results exhibit stronger pooling across diameters. The key observation is that in both cases, the posteriors remain diameter-specific, while incorporating shared information across the diameters of each \textsc{emb} model.

\begin{figure}[t]
\centering
\includegraphics[width=\textwidth]{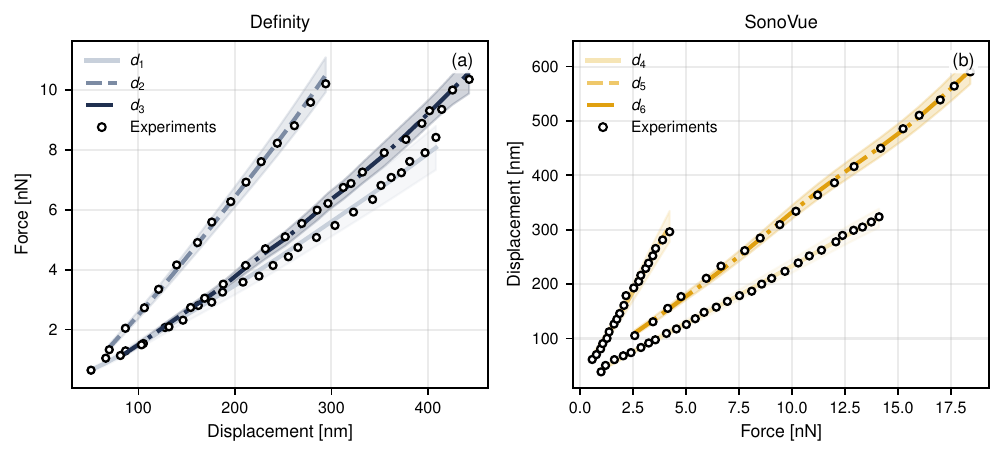}
\caption{\textbf{Reduced-model posterior predictive bands.} Panels (a) and (b) show the Definity (ink) and SonoVue (gold) models, respectively, with all three diameters overlaid in each case. For each diameter, the lines denote the reduced hierarchical posterior median response, the shaded region denotes the 90\% posterior predictive band, and the circles denote the experimental points.}
\label{fig:reduced_bands_all}
\end{figure}
We assess predictive agreement in the reduced-model posterior by propagating, for each bubble $i$, samples from the reduced hierarchical posterior $P(\tilde\theta_{d_i})$ together with the inferred associated noise parameter $\sigma$ through the \textsc{dnn} surrogate $\mathcal{S}_{d_i}^0$ on the experimental reference grid (Methods). Fig.~\ref{fig:reduced_bands_all} reports the resulting reduced-models posterior predictive bands together with the reduced median response and the experimental points. The reduced posteriors predictive bands reproduce the dominant experimental trends with good fidelity, with coverage ranges from 93.1\% to 100.0\%. These results therefore show that the reduced models are in good agreement with the experimental curves once both posterior parameter uncertainty and inferred observation noise are propagated through the surrogates.

\label{sec:res:map_confirmation}

\begin{figure}[t]
\centering
\includegraphics[width=\textwidth]{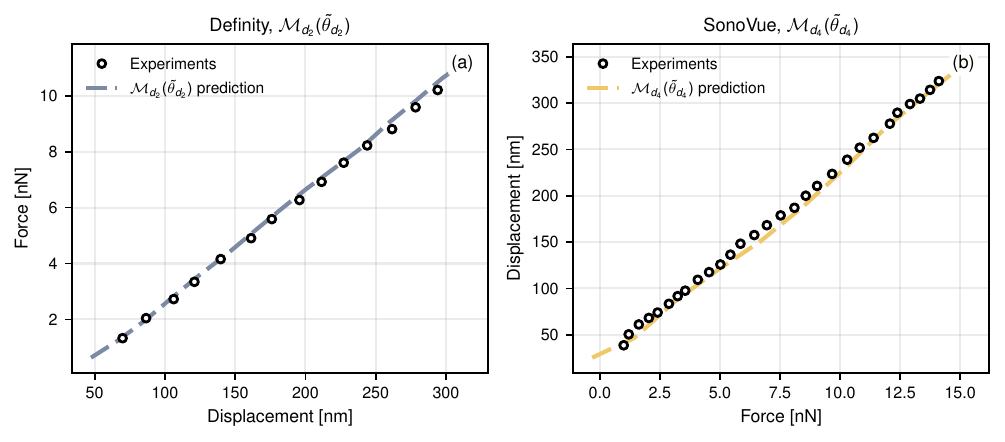}
\caption{\textbf{Representative reduced-model \textsc{map} confirmation against direct \textsc{dpd} simulations}. Panels (a) and (b) depict $\mathcal{M}_{d_2}(\tilde\theta_{d_2})$ (ink) and $\mathcal{M}_{d_4}(\tilde\theta_{d_4})$ (gold), respectively. In each panel, the circles denote the experimental data, and the dashed line denotes a sparse 15-point direct \textsc{dpd} response evaluated at the reduced-model \textsc{map} parameters.}
\label{fig:map_confirmation_reduced}
\end{figure}

To confirm that the surrogate-based reduced calibrations remain faithful to the original simulator at representative reduced-model \textsc{map} parameter sets, we compare reduced-model \textsc{map} curves from \textsc{dnn}-guided inferences with direct \textsc{dpd} simulations (See Fig.~\ref{fig:workflow}, block \circled{4}). Because propagating full posterior uncertainty directly through the \textsc{dpd} simulator would require a prohibitively large number of expensive \textsc{dpd} simulations, we restrict the direct simulator-side confirmation in the present paper to sparse \textsc{map}-level overlays. For both Definity and SonoVue models, the direct reduced-model \textsc{map} \textsc{dpd} responses compare well with the experimental data. Fig.~\ref{fig:map_confirmation_reduced} displays this comparison for $\mathcal{M}_{d_2}(\tilde\theta_{d_2})$ and $\mathcal{M}_{d_4}(\tilde\theta_{d_4})$. For these two direct checks, the relative $L^2$ discrepancy over the overlapping force range between the direct \textsc{dpd} response and the reference curve is 2.50\% for $\mathcal{M}_{d_2}$ and 3.36\% for $\mathcal{M}_{d_4}$. This provides a representative direct simulator-side consistency check at reduced-model \textsc{map} points.

\label{sec:res:reduced_model}
  
\begin{figure}[t]
\centering
\includegraphics[width=\textwidth]{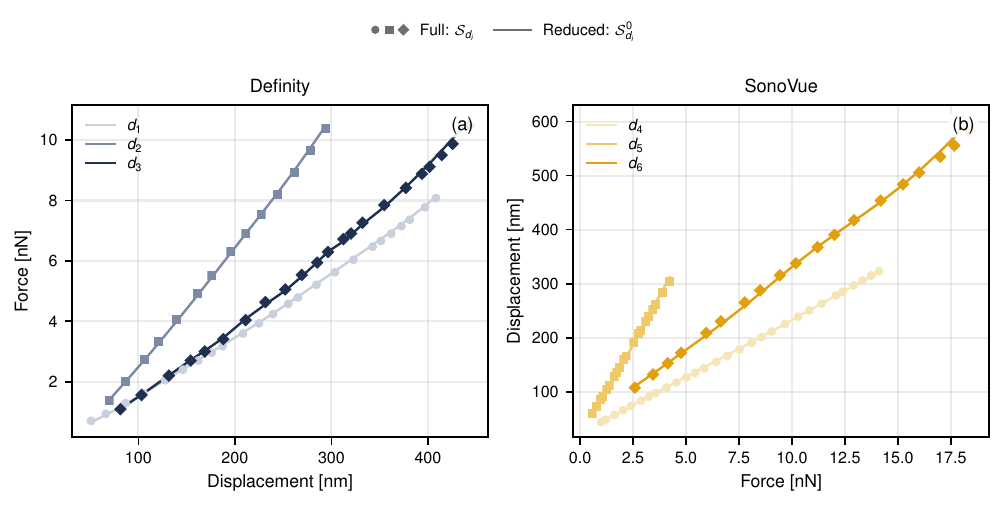}
\caption{\textbf{Final full- versus reduced-model surrogate sanity check.} Panels (a) and (b) show Definity (ink) and SonoVue (gold), respectively. In both panels, the solid lines denote the reduced-model \textsc{map} surrogate responses, while the symbols denote the full-model \textsc{map} surrogate responses. The figure provides a direct \textsc{map}-level overlay check, showing that the full and reduced workflows are close in the present quasi-static regime.}
\label{fig:full_vs_reduced_map_overlay}
\end{figure}

After carrying the reduced model through the main inference workflows, we rerun the full force field inference once for the Definity and SonoVue models to check whether the weakly informed nonlinear terms materially change the fitted response. Using the surrogate forward models with the \textsc{map} parameter sets, we quantify the discrepancy between the full and reduced \textsc{map} responses via the relative $L^2$ difference, $100\lVert y_{\mathrm{full}}^{\mathrm{MAP}}-y_{\mathrm{red}}^{\mathrm{MAP}}\rVert_2/\lVert y_{\mathrm{full}}^{\mathrm{MAP}}\rVert_2$.
The worst-case discrepancy across diameters remains below 1.2\% for Definity and below 1.9\% for SonoVue. Fig.~\ref{fig:full_vs_reduced_map_overlay} compares the full- and the reduced-model \textsc{map} surrogates in real units, showing that the two workflows are close at the level of the fitted response. This agreement is further supported by the \textsc{crps} \cite{matheson1976scoring,hersbach2000decomposition}, which measures how closely the full posterior predictive distribution matches the observed data while rewarding both calibration and sharpness, and shows comparable performance of the full- and reduced-models in terms of coverage and statistical consistency with the reference data (Supplementary Table~\ref{tab:crps_trusted}). Combined with the \textsc{map} overlays, these \textsc{crps} values show that the reduced workflow preserves essentially the same probabilistic predictive quality as the full workflow for both Definity and SonoVue models. We therefore use the full workflow here only as a confirmatory safety check, not as a co-equal narrative track or as the basis for a formal model-selection claim. The practical conclusion is that, in the present quasi-static regime, the nonlinear elastic terms are weakly informed and the reduced model captures the dominant behavior that the data can support.

For both Definity and SonoVue \textsc{emb} models, \textsc{dnn} surrogates reproduced the \textsc{dpd} response curves with low validation error (Table~\ref{tab:surrogate_validation}), allowing tractable Bayesian calibration over thousands of forward-model evaluations (Fig.~\ref{fig:workflow}, block \circled{1}). The global sensitivity results (Fig.~\ref{fig:sensitivity}) showed that the response in the explored regime was governed primarily by $k_a$ and $k_b$, while the nonlinear coefficients $(b_1,b_2,a_3,a_4)$ remained weakly informed (Fig.~\ref{fig:workflow}, block \circled{2}). Carrying only the reduced model after the sensitivity section (Fig.~\ref{fig:workflow}, block \circled{3}) yielded well-defined reduced independent posteriors (Fig.~\ref{fig:phase1_reduced_representative}) and, after hierarchical pooling, reduced posteriors that remained stable while incorporating pooled information across diameters for each \textsc{emb} model (Fig.~\ref{fig:phase1_vs_phase3b_reduced_summary}). The resulting reduced posterior predictive bands reproduced the dominant experimental trends with good fidelity (Fig.~\ref{fig:reduced_bands_all}). Direct \textsc{dpd} simulations at reduced-model \textsc{map} parameter sets then provided simulator-side confirmation (Fig.~\ref{fig:map_confirmation_reduced}) and allowed us to obtain data-informed \textsc{emb} models that are quantitatively calibrated to the experimental data while remaining consistent with the underlying \textsc{dpd} physics (Fig.~\ref{fig:workflow}, block \circled{4}). Finally, rerunning the full workflow once at the end showed that the worst-case full-vs-reduced \textsc{map} discrepancy was low (Fig.~\ref{fig:full_vs_reduced_map_overlay}), while the corresponding curve-averaged posterior-predictive \textsc{crps} remained close to the full-model values (Supplementary Table~\ref{tab:crps_trusted}). These results support the reduced force field as the effective calibration model used in this paper, with the full model retained only as a final sanity check. The final reduced hierarchical \textsc{map} values of the dominant elastic parameters are summarized in Table~\ref{tab:map_parameters_reduced}.

\begin{table}[H] 
\centering
\caption{\textbf{Reduced hierarchical \textsc{map} values of the dominant elastic parameters across different types and diameters (\textsc{dpd} units).}}
\label{tab:map_parameters_reduced}
{
\footnotesize
\setlength{\tabcolsep}{6pt}
\begin{tabular}{l|c|c|c}
\hline
\textbf{Type} & \textbf{Diameter} & $k_a$ & $k_b$ \\
\hline
 & $d_1$ (2.1 $\mu$m) & 7,843.2 & 192.1 \\
Definity & $d_2$ (2.9 $\mu$m) & 18,797.6 & 915.8 \\
 & $d_3$ (3.0 $\mu$m) & 13,797.8 & 432.5 \\
\hline
 & $d_4$ (3.2 $\mu$m) & 15,009.1 & 12,569.4 \\
SonoVue & $d_5$ (3.4 $\mu$m) & 1,656.0 & 12,739.3 \\
 & $d_6$ (5.8 $\mu$m) & 20,766.9 & 12,635.1 \\
\hline
\end{tabular}
}
\end{table}

\section{Discussion}
\label{sec:discussion}

The central result of this study is that we develop and apply a surrogate-accelerated Bayesian workflow to a high-fidelity numerical \textsc{emb} model with quantified uncertainty, and show that, in the present quasi-static regime, the data constrain $k_a$ and $k_b$ and support a reduced force field description. Quantitatively, grouped-holdout surrogate errors remain below 2.1\% for compression and 2.6\% for indentation, and the worst-case full-vs-reduced \textsc{map} discrepancy remains below 1.18\% for Definity and 1.82\% for SonoVue at
the surrogate \textsc{map} level. The present contribution is therefore the quasi-static calibration and identifiability analysis itself, not an acoustic-validation study. In both \textsc{emb} types, the Results section indicates a consistent pattern: the stretching and bending responses dominate the observable force--displacement behavior, whereas the nonlinear coefficients remain only weakly constrained (Sections~\ref{sec:res:sensitivity} and \ref{sec:res:phase1}). That is why we resort to a reduced model immediately after the sensitivity analysis and retain the full model only for the final sanity check in Section~\ref{sec:res:reduced_model}. The conclusion of the reduced model is therefore not based on generic visual agreement alone, but on one explicit chain of evidence: dominant $k_a/k_b$ sensitivities, stable reduced posteriors before and after hierarchical pooling, close full-vs-reduced \textsc{map} overlays, and close reduced-vs-full posterior-predictive \textsc{crps}. This distinction between identifiable and weakly identifiable parameters is important, because it distinguishes uncertainty intrinsic to the data from uncertainty introduced by the inference process itself.

The comparison between independent and hierarchical inference clarifies that the hierarchical construction serves primarily as a regularization device. Independent inference is useful as a baseline because it unveils the information content of each dataset in isolation (Section~\ref{sec:res:phase1}). Hierarchical inference, on the contrary, does not generate information where none exists; instead, it regularizes diameter-specific reduced-model posteriors through population-level structure learned in the hyperparameter inference phase (Section~\ref{sec:res:phase2}). This is most helpful for parameters that are only weakly informed by a single curve because those directions are the most susceptible to broad posteriors, correlations, and sampling noise. The Gaussian form used for the hierarchical conditionals should therefore be understood as a pragmatic partial-pooling assumption, rather than as evidence that the true across-diameter parameter distributions are exactly normal. For the dominant elastic parameters, the effect of hierarchy is more modest in compression than in indentation. The compression curves already provide substantial information about the leading stiffness scales, so hierarchy mainly stabilizes the final posteriors rather than changing their interpretation. For indentation, the updated reduced summaries still indicate visible partial-pooling effects across all three diameters. This behavior is consistent with previous hierarchical calibration studies in biomechanics, where partial pooling improves robustness without enforcing exact parameter equality across datasets~\citep{economides2021hierarchical}. In that limited sense, hierarchical pooling may also help generate regularized parameter sets for newly observed \textsc{emb}s from the same shell-chemistry class. In the present problem, that regularization is useful precisely because it acts on the reduced parameterization that the data can support, rather than attempting to recover the weakly informed nonlinear directions. We do not interpret this as a generic transfer claim across chemistries or observables.

Our modeling framework is inspired by recent data-driven red blood cell models \cite{walchli2020load,economides2021hierarchical,amoudruz2022simulations,amoudruz2023stress,amoudruz2024volume,amoudruz2025scalable}, which demonstrate that nonlinear elastic coefficients, particularly the shear-hardening parameter $b_2$, play a significant role under large stretching deformations. In particular, these studies indicate that $b_2$ becomes influential in the high-force regime, where strain-hardening effects dominate the mechanical response. Motivated by these findings, we systematically investigate the role of nonlinear elastic parameters $(b_1,b_2,a_3,a_4)$ in deformation regimes relevant to \textsc{afm} experiments, including both indentation and compression, for which quantitative force--displacement data are available. This allows us to assess whether such higher-order contributions are required to explain experimental observations in moderate deformation regimes, where their importance remains unclear.

The formulation of the reduced model constitutes an important component of the present work. When the nonlinear terms are set to zero, the dominant fitted responses remain close to those of the full workflow at the level of \textsc{map} predictions, and the final full-model rerun confirms that the main response envelopes do not change materially in the present regime (Section~\ref{sec:res:reduced_model}). The bands shown in the reduced-model workflow are pointwise posterior predictive intervals under the current heteroscedastic Gaussian likelihood, and their reduced-model coverage ranges from 93.1\% to 100.0\% across the six datasets. Because the current workflow still lacks an explicit surrogate-error model, we interpret this high coverage as evidence of good agreement under the assumed statistical model rather than as a complete validation of every uncertainty source. At the same time, because the likelihood uses $s_{ij}=\sigma_i\hat{y}_{ij}$, the uncertainty shrinks as the predicted response approaches zero. In the low-load regime near first contact, where alignment ambiguity matters most, this likely understates uncertainty and should be expected to affect $d_0$ more strongly than the dominant elastic parameters. The reduced model also yields simpler and more stable posteriors for the parameters constrained by the experiments. Taken together, these results show that the present quasi-static compression and indentation datasets support the reduced model for calibration purposes as a parsimonious, better-conditioned description, with the full model retained only as a confirmatory safety check. This is exactly the pattern expected when additional degrees of freedom are formally available in the model but are not supported by the data.

This result should not be interpreted as evidence that the nonlinear interactions are unphysical or universally negligible. A more careful interpretation is that, within the deformation regime probed here, they do not produce an identifiable predictive signal beyond what is already captured by the dominant elastic terms and the calibration parameter $d_0$. In practice, this means that the current experiments are sufficient for calibrating an effective reduced model, but not for validating the full nonlinear force field. Without an explicit comparison based on a widely applicable information criterion, leave-one-out cross-validation, or evidence / marginal likelihood, we do not present this as a formal model-selection verdict; those comparisons are deferred to future work. Instead, the simpler model is preferable for the present task because it achieves nearly the same dominant predictive behavior with fewer weakly constrained parameters and better-conditioned posteriors.

In force--deformation measurements of \textsc{emb}, uncertainty in the precise location of first contact introduces a systematic ambiguity in the deformation coordinate. In \textsc{afm} studies, this issue is handled implicitly through data-processing steps such as translating the force--distance curve to define zero separation at an assumed contact point or excluding the initial nonlinear region before fitting. Although effective in practice, these procedures are equivalent to introducing an unreported displacement offset. The explicit inference of the displacement offset $d_0$ is another important outcome of the study. In both the independent and hierarchical stages, allowing a small curve-alignment correction removes a systematic mismatch near first contact and improves agreement in the low-load regime. In the present framework, $d_0$ is hierarchically pooled within each bubble type as a pragmatic regularization choice, whose impact on the inferred posteriors could be worth assessing in future work. It is a calibration quantity introduced to compensate for the initial bubble loading state and contact or alignment details in the published reference curves that the present \textsc{dpd} model does not attempt to represent. It accounts for a real experimental ambiguity associated with contact-point determination and alignment, preventing that ambiguity from being spuriously mapped onto shell stiffness. Methodologically, $d_0$ illustrates a broader point about Bayesian calibration of experimental data: nuisance parameters that encode measurement protocol effects can be essential for obtaining physically interpretable posteriors. If such effects are ignored, the model may compensate by distorting constitutive parameters, which then appear artificially well constrained. Here, inferring $d_0$ makes the remaining mechanical parameters more credible precisely because it acknowledges that the experimental observable is not a perfectly aligned representation of the underlying shell response.

Treating Definity and SonoVue \textsc{emb}s separately, but within a unified methodological framework highlights both shared structure and size-specific behavior. Across all datasets, the dominant information concerns the leading elastic response, while the nonlinear coefficients remain weakly informative. This consistency indicates that the weak identifiability of the nonlinear terms is not an artifact. The sensitivity analysis indicates that the relative contributions of stretching and bending depend on both \textsc{emb} type and diameter. Both \textsc{emb} types exhibit a distinct balance between stretching and bending, modulated by size. Because shell composition and diameter change together, the comparison provides insight into shared trends and broad differences, but does not isolate their individual effects. The present paper analyzes the two \textsc{emb} types in parallel rather than in a single joint posterior, which is appropriate for establishing the workflow, while a multimodal hierarchical calibration remains a useful longer-term methodological extension.

Several limitations should be considered when interpreting the inferred parameters. First, the forward model is still an effective mesoscale description. Even with Bayesian calibration, the inferred force field parameters remain model-dependent summaries of shell mechanics instead of direct measurements of a unique underlying material law. This point is especially important because the present study reanalyzes only six reprocessed published curves, one representative curve per diameter, rather than a full cohort of bubbles within each formulation. The resulting posteriors therefore describe what can be learned from those six curves under the present model, not cohort-level material-property distributions for Definity or SonoVue. Second, the current likelihood and propagation workflow treats each \textsc{emb} type with a separate surrogate (Section~\ref{sec:meth:bayes}). This is a pragmatic choice reflecting the computational pipeline's implementation, but it means that the statistical model is not yet fully unified across observables. A future version of the framework should make that symmetry more explicit, especially if compression and indentation are to be fused into one joint calibration. Third, the conclusions are tied to the current experimental regime. The data are quasi-static and relatively low dimensional, over a limited loading range for six diameter-specific datasets. Under these conditions, it is unsurprising that the leading elastic scales are identifiable, while the higher-order nonlinear terms are not. If the scientific goal shifts toward validating the full force field, richer observables will be necessary. Time-dependent loading, oscillatory forcing, rupture-adjacent deformations, or independent measurements of the shell microstructure may all provide complementary information that is absent from the current curves. Finally, the workflow relies on surrogates trained on finite simulation databases. The validation errors reported in Table~\ref{tab:surrogate_validation} are low enough to make the inference tractable and credible, but the surrogate error is not zero. The practical implication is that the posterior should be interpreted as conditional on both the \textsc{dpd} model and its learned surrogate representation. In the current workflow, that surrogate error is effectively treated as zero during inference and propagation, so only parameter uncertainty and the inferred observation noise enter the displayed posterior predictive bands. Future extensions could address this limitation by introducing an explicit surrogate-uncertainty model, for example through Bayesian \textsc{dnn}s or other probabilistic surrogates, so that surrogate uncertainty can be propagated alongside the current likelihood-based noise model. This would allow surrogate uncertainty to be distinguished from observational noise, yielding more faithful posterior credible intervals and more reliable \textsc{uq}. A related limitation is that the paper does not report posterior predictive bands computed directly with the \textsc{dpd} simulator itself: propagating a large posterior ensemble through \textsc{dpd} would be prohibitively expensive with the current simulator cost. Likewise, the direct \textsc{dpd} overlays reported in Results are representative \textsc{map}-level checks only; they are useful simulator-side confirmations at selected parameter points, but not exhaustive validations over all diameters or over full posterior regions. This caveat is standard in scientific machine learning~\citep{raissi2019physics,liu2021uncertainty}, but it remains important when drawing physical conclusions from a calibrated surrogate pipeline.

In conclusion, we developed data-informed \textsc{dpd} models that quantify the stretching stiffness and bending modulus of two types of lipid-\textsc{emb}s using a surrogate-accelerated Bayesian inference framework. Our models accounted for variations in \textsc{emb} size, elastic parameters, displacement offset, and measurement noise. Model parameters were calibrated against three datasets for each \textsc{emb} type, based on force spectroscopy compression and indentation measurements. The use of \textsc{dnn} surrogates enabled efficient posterior sampling while maintaining high fidelity to the underlying \textsc{dpd} simulations. Parameter uncertainties were propagated through forward predictions to assess model robustness and transferability.
In addition to these uncertainties, the assumption of a constant membrane thickness introduces an additional source of uncertainty in the inferred elastic parameters. As noted in \textsc{afm} studies of phospholipid microbubbles, apparent variations in Young's modulus with \textsc{emb} size may arise from underlying variations in shell thickness. In this case, the inferred elastic modulus limits the ability to disentangle geometry from material response. To address this identifiability limitation, a natural long-term extension of the current framework would be to treat the membrane thickness as an additional parameter in the inference. The next step, beyond the scope of the present quasi-static study, will be to test how the calibrated models perform in acoustic simulations. That setting provides dynamic observables absent from the current quasi-static datasets and thus offers a natural way to assess whether the reduced force field remains sufficient beyond compression and indentation. Such studies would also show how calibration uncertainty in the mechanical parameters propagates to acoustically relevant predictions. If direct \textsc{dpd} uncertainty bands become necessary, one possible route would be to use an Unscented Transform built around a small set of carefully selected \textsc{dpd} evaluations, so that simulator-side uncertainty bands can be approximated without brute-force posterior propagation. Joint multimodal inference remains an interesting methodological extension, but the more direct scientific payoff in future work would be to carry the calibrated posterior distributions into acoustically relevant simulations and compare the resulting behavior across model choices. More broadly, the presented framework shows that rigorous \textsc{uq} for particle-based soft-matter models is feasible once high-fidelity simulations are coupled with surrogates and hierarchical Bayesian inference. This combination enables parsimony, calibration, and identifiability questions that would be computationally inaccessible with direct \textsc{dpd} sampling alone. Even though the present study ultimately favored a reduced force field in this experimental regime, this still represents a meaningful scientific outcome, identifying the level of model complexity supported by the current data and establishing clear targets for additional dynamic data and simulations required to justify more complex models. Our tailor-made force spectroscopy data-informed \textsc{emb} models provide a basis for large-scale acoustic simulations to determine key ultrasound parameters \cite{lah2025open}, such as intensity, pulse duration, and frequency for \textsc{usdg}. The methodology can be further extended to model development of ligand-functionalized \textsc{emb}s or encapsulated gas vesicles \cite{maresca2018biomolecular,salahshoor2022geometric,dutka2023structure,smith2025ultrafast,ntarakas2025dissipative}, and contrast agents for other imaging techniques including echocardiography \cite{kaufmann2007contrast}, magnetic resonance imaging, positron emission tomography, diffraction-enhanced X-ray imaging \cite{kogan2010microbubbles} and X-ray dark-field imaging \cite{velroyen2013microbubbles}.

\newpage
\section{Methods}
\label{sec:methods}

\begin{figure}[t]
\centering
\includegraphics[width=1.\textwidth]{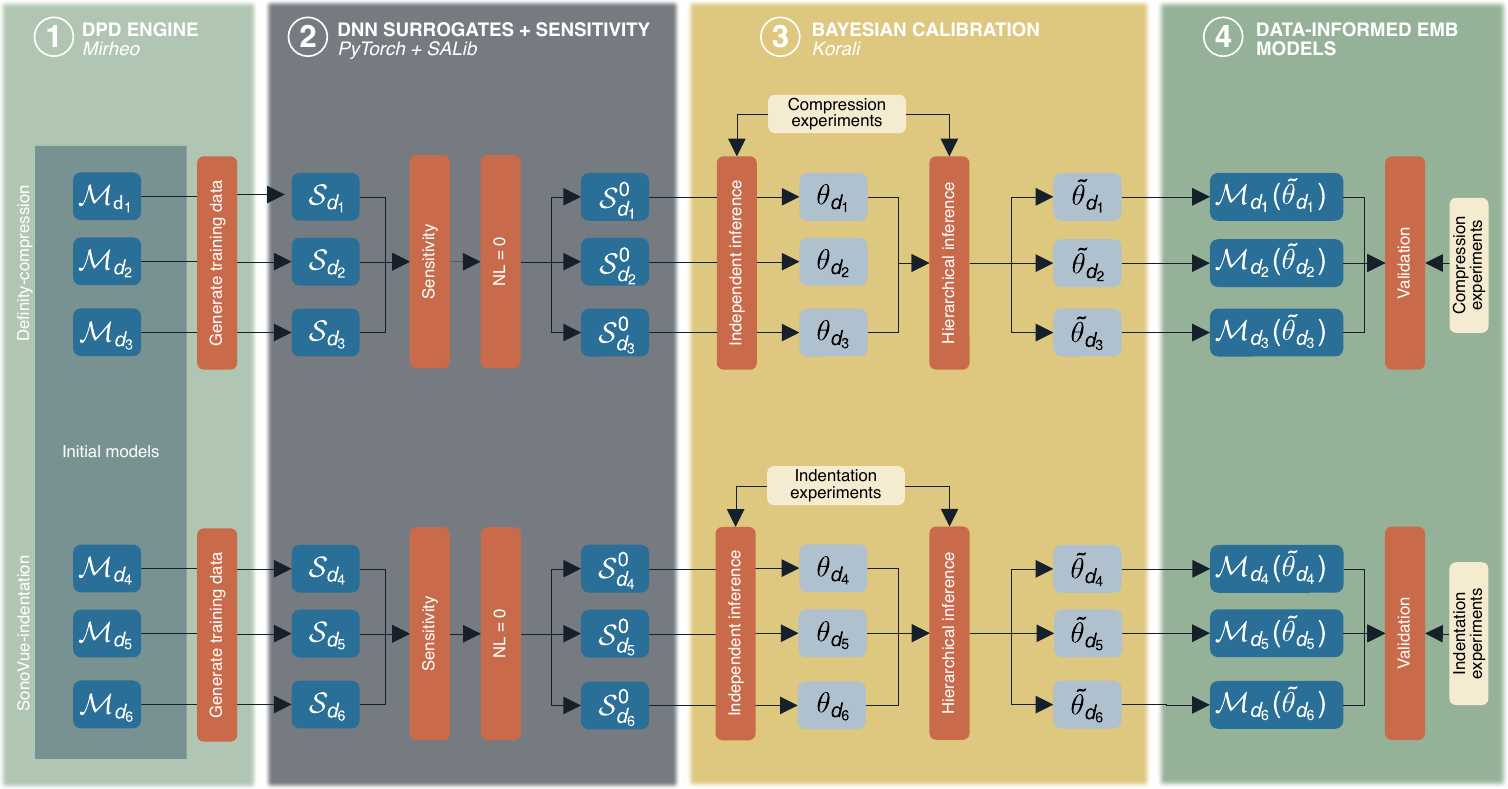}
\caption{\textbf{Workflow diagram for the hierarchical Bayesian inference of data-informed \textsc{emb} models.} $\mathcal{M}_{d_1,d_2,d_3}$ denote models of Definity bubbles, while $\mathcal{M}_{d_4,d_5,d_6}$ denote SonoVue bubbles. Throughout the workflow, the index $d_i$ labels the bubble diameter, with $(d_i)_{i=1,2,3,4,5,6}$ corresponding to $2.1$, $2.9$, $3.0$, $3.2$, $3.4$, and $5.8\,\mu$m, respectively. For each bubble $i$, the \textsc{dpd} model $\mathcal{M}_{d_i}$ generates training data used to construct a \textsc{dnn} surrogate $S_{d_i}$, which is subsequently reduced to $S_{d_i}^{0}$ through sensitivity analysis and suppression of nonlinear terms. Each reduced surrogate is then combined with the corresponding compression (top row) or indentation (bottom row) experiments in a Bayesian calibration workflow consisting of an independent inference step, yielding model parameters $\theta_{d_i}$, followed by a hierarchical inference step, yielding regularized parameters $\tilde{\theta}_{d_i}$. The resulting calibrated model $\mathcal{M}_{d_i}(\tilde{\theta}_{d_i})$ is finally validated against the corresponding experimental data.}
\label{fig:workflow}
\end{figure}

\textbf{\\\textsc{dpd} simulations}
\label{sec:meth:dpd}

\noindent All mesoscale simulations were performed using \textsc{dpd}, a particle-based method well suited for modeling soft matter systems with explicit hydrodynamics~\citep{groot1997dissipative}. In \textsc{dpd}, each particle represents a coarse-grained cluster of molecules and evolves according to Newton's equations of motion. The total force acting on the particle $i$ is given by
\begin{equation}
\mathbf{F}_i = \sum_{j \neq i} \left(\mathbf{F}_{ij}^{\mathrm{C}} + \mathbf{F}_{ij}^{\mathrm{D}} + \mathbf{F}_{ij}^{\mathrm{R}} \right) + \mathbf{F}_i^{\mathrm{shell}},
\end{equation}
where $\mathbf{F}_{ij}^{\mathrm{C}}$, $\mathbf{F}_{ij}^{\mathrm{D}}$, and $\mathbf{F}_{ij}^{\mathrm{R}}$ are the conservative, dissipative, and random pairwise forces, respectively, and $\mathbf{F}_i^{\mathrm{shell}}$ denotes the elastic forces arising from the encapsulating microbubble shell.

The conservative force between particles $i$ and $j$ is defined as
\begin{equation}
\mathbf{F}_{ij}^{\mathrm{C}} =
\begin{cases}
a_{ij} \left(1 - r_{ij}/r_c \right) \hat{\mathbf{r}}_{ij}, & r_{ij} < r_c, \\
0, & r_{ij} \geq r_c,
\end{cases}
\end{equation}
where $r_{ij}$ is the interparticle distance, $\hat{\mathbf{r}}_{ij}$ the corresponding unit vector, and $r_c$ the cutoff radius. Dissipative and random forces are given by
\begin{align}
\mathbf{F}_{ij}^{\mathrm{D}} &= -\gamma \, \omega^{\mathrm{D}}(r_{ij}) 
\left(\hat{\mathbf{r}}_{ij} \cdot \mathbf{v}_{ij} \right) \hat{\mathbf{r}}_{ij}, \\
\mathbf{F}_{ij}^{\mathrm{R}} &= \sigma \, \omega^{\mathrm{R}}(r_{ij}) \, \xi_{ij} \, \Delta t^{-1/2} \hat{\mathbf{r}}_{ij},
\end{align}
where $\mathbf{v}_{ij}$ is the relative velocity, $\xi_{ij}$ is a zero-mean unit-variance Gaussian random variable, and $\Delta t$ is the integration time step. The \textsc{dpd} settings that were used throughout were $\gamma = 3.5$, $\sigma^2 = 2 \gamma k_B T$, $k_B T = 1$, and $\omega^{\mathrm{D}}(r) = (1 - r/r_c)^2$ for $r < r_c$.

The encapsulating microbubble shell was modeled as a triangulated elastic surface immersed in a \textsc{dpd} solvent. Each vertex of the triangulation corresponds to a \textsc{dpd} particle, and elastic forces are obtained as derivatives of a discrete shell energy. Using expressions adopted from our previous work \cite{ntarakas2025dissipative}, the active shell energy in the present \textsc{emb} workflow is
\begin{equation}
E_{\mathrm{shell}} = E_{\mathrm{stretch}} + E_{\mathrm{bend}}.
\end{equation}
In-plane elastic deformations are described by
\begin{equation} \label{stretch_ene}
  E_{\mathrm{stretch}} = \frac{k_a}{2} \sum_{t=1}^{N_t} A_t^0[\alpha_t^2 + a_3 \alpha_t^3 + a_4 \alpha_t^4] + \mu \sum_{t=1}^{N_t} A_t^0[\beta_t + b_1 \alpha_t \beta_t + b_2 \beta_t^2],
\end{equation}
where $\alpha_t$ and $\beta_t$ are the area and shear strain invariants of the triangle $t$ and $A_t^0$ is the area of triangle $t$. This expression also includes nonlinear terms (coefficients $a_3, a_4, b_1$, $b_2$). Bending resistance is modeled via a dihedral-angle potential,
\begin{equation} \label{bend_ene}
  E_{\mathrm{bend}} = k_b\sum_{\langle i,j\rangle} \left[1-\cos(\theta_{ij}-\theta_{ij}^0)\right],
\end{equation}
with $\theta_{ij}$ and $\theta_{ij}^0$ the actual and spontaneous angles between two adjacent triangles. We report the inferred in-plane stretching scale directly as $k_a$. The modeling choice for the in-plane coefficient $\mu$ is deliberate.  We infer $k_b$ independently, as the stretching--bending relationhip is model-dependent and less reliable for coated microbubbles. In contrast, we retain an isotropic in-plane law \cite{lim2002stomatocyte,Lim_H_W2009-hl}, since the data constrain only the dominant stiffness and do not support identification of a separate shear modulus. Accordingly, $\mu$ is not inferred independently and is related to $k_a$ through 
\begin{equation}
\mu = \frac{1-\nu}{1+\nu}\,k_a,
\end{equation}
which gives $\mu = k_a/3$ for the fixed Poisson ratio $\nu=0.5$ used throughout. The higher-order nonlinear elastic parameters $(b_1, b_2, a_3, a_4)$ are introduced to account for the large strains encountered in strongly deformed red blood cell morphologies, where linear elasticity fails, and these terms capture the strain-hardening response of the spectrin cytoskeleton, resulting in a marked increase in membrane stiffness and deformation energy at large strain \cite{lim2002stomatocyte,Lim_H_W2009-hl,bozic2012role}.

Both the surrounding liquid and the encapsulated gas were modeled explicitly using \textsc{dpd} particles. No-through boundary conditions at the membrane were enforced using a bounce-back mechanism~\citep{revenga1999boundary,fedosov2010multiscale}, while interfacial slip conditions were controlled by tuning the dissipative coupling parameters between shell, water, and gas particles. The treatment of the gas phase, including pressure compensation and ideal-gas behavior, follows exactly the methodology described in~\cite{ntarakas2025dissipative}. All simulations were performed using the \textsc{gpu}-accelerated \textsc{dpd} framework \textsc{mirheo} \cite{alexeev2020mirheo}, using reduced \textsc{dpd} units with $k_B T = 1$, and number density $n_d = 3$.
Physical units were recovered through a characteristic length scale of $250~\mathrm{nm}$, thermal energy at room temperature, and mass scaling based on the density of water. To ensure computational feasibility while preserving the relevant elastic physics, all elastic moduli of the shell were uniformly downscaled by a factor $f_{scale} = 0.0074$ \cite{ntarakas2025dissipative}, which preserves the F\"oppl--von K\'arm\'an number \cite{paulose2012fluctuating} and thus the equilibrium and deformation behavior of the shell. Physical forces were recovered by rescaling during post-processing \cite{ntarakas2025dissipative}. The mechanical response of the shell was characterized by extracting force–displacement relations through two distinct simulation protocols: indentation and compression.

\begin{figure}
    \centering
    \includegraphics[width=1.0\linewidth]{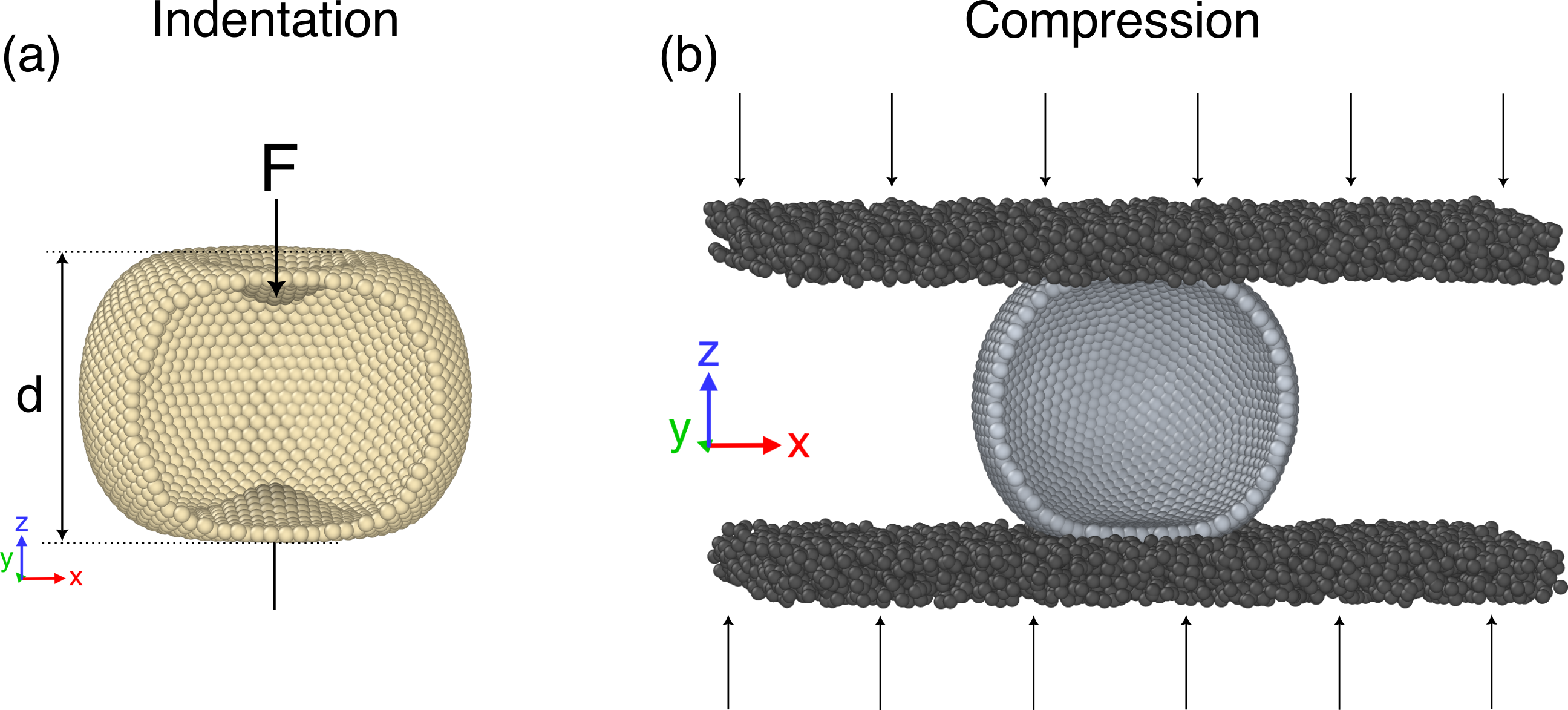}
    \caption{\textbf{A schematic representation of the \textsc{dpd} simulations implemented in this work.} (a) Indentation simulations are performed on a SonoVue microbubble (gold) by applying a pair of oppositely equal forces to a small fraction of vertices at the top and bottom of a microbubble. (b) The plate compression is implemented by displacing a set of frozen \textsc{dpd} particles, representing a large cantilever used in the compression of a Definity microbubble (ink).}
    \label{fig:dpd_scheme}
\end{figure}

Indentation simulations mimic \textsc{afm} indentation experiments by applying equal and opposite external forces to sets of shell vertices located near the top and bottom poles of the microbubble, see Fig.~\ref{fig:dpd_scheme}(a). The equatorial vertices were constrained in the axial direction to suppress rigid-body motion. For each applied force level, the system was equilibrated quasi-statically, and the resulting axial displacement of the shell was recorded. Force--displacement curves were obtained by prescribing the force and measuring the displacement. Parallel-plate compression simulations were implemented using two rigid plates represented by a dense collection of frozen particles, see Fig.~\ref{fig:dpd_scheme}(b). The plates were pinned to move only along the axial direction and displaced quasi-statically toward the microbubble. After equilibration, the reaction force exerted by the shell on the plates was recorded. In this case, force--displacement curves were obtained by prescribing displacement and measuring the resulting reaction force.

\newpage
\textbf{\\Deep neural network (\textsc{dnn}) surrogates}
\label{sec:methods_surrogates}

\noindent Direct evaluation of the mechanical response of an \textsc{emb} with \textsc{dpd} simulations is computationally expensive, making it impractical as the forward model in large-population Bayesian inference runs. Therefore, we replaced the \textsc{dpd} solver and corresponding models $(\mathcal{M}_{d_i})_{i \in \llbracket 1,6 \rrbracket}$ with \textsc{dnn} surrogate models $(\mathcal{S}_{d_i})_{i \in \llbracket 1,6 \rrbracket}$ trained on simulation data, and used these surrogates as fast emulators during inference and uncertainty propagation (See Fig.~\ref{fig:workflow} for details). In all cases, the surrogate was trained to approximate \textsc{dpd} outputs for a fixed \textsc{emb} model and a fixed \textsc{emb} diameter. Six surrogates were constructed, one for each of the three diameters within both \textsc{emb} models. All surrogate inputs were written directly in terms of $(k_a,k_b,\ldots)$, consistent with the parameterization used throughout the Results section.
For Definity \textsc{emb}s, the surrogates $S_{d_1}, S_{d_2}$ and $S_{d_3}$ approximated the forward map
\begin{equation}
  \bigl(k_a, k_b, b_1, b_2, a_3, a_4, d\bigr) \;\mapsto\; F,
\end{equation}
In other words, it predicted the compressive force $F$ at displacement $d$ for a given set of material parameters.
For SonoVue \textsc{emb}s, the surrogates $S_{d_4}, S_{d_5}$ and $S_{d_6}$ approximated the map
\begin{equation}
  \bigl(k_a, k_b, b_1, b_2, a_3, a_4, F\bigr) \;\mapsto\; d,
\end{equation}
Namely, it predicted the indentation displacement $d$ at applied force $F$. 
We trained a separate surrogate per diameter and stored the best-performing model for each diameter for use in the inference pipeline. Implementation details, including how the best model was selected, are summarized in Supplementary Section~\ref{app:surrogate}. 

\textbf{\\Bayesian inference framework}
\label{sec:meth:bayes}

\noindent We calibrated the \textsc{emb} force-field parameters with \textsc{tmcmc}~\cite{ching2007transitional} as implemented in \textsc{korali}~\cite{martin2022korali}. The full and reduced workflows share the same inferential structure; unless otherwise stated, the reported results correspond to the reduced model, in which $(b_1,b_2,a_3,a_4) = (0,0,0,0)$, while the full model is reported as a confirmatory analysis. For each diameter-specific reference curve $i$, the reference dataset is a set of $N_i$ paired observations $\mathcal{D}_i = \{(x_{ij}, y_{ij})\}_{j=1}^{N_i}$. For compression (resp. indentation) experiments, $x_{ij}$ denotes displacement (resp. force) and $y_{ij}$ denotes force (resp. displacement). The diameter-specific parameter vector is $\Theta_{d_i} = \bigl(k_a, k_b, b_1, b_2, a_3, a_4, d_0\bigr)_{d_i}$, where $(k_a,k_b,b_1,b_2,a_3,a_4)$ are the constitutive parameters and $d_0$ is a calibration offset introduced because the initial bubble-cantilever contact establishment is outside the scope of the present \textsc{dpd} model. The observation-noise amplitude is denoted separately by $\sigma_{d_i}$ and is shared across all points of a given curve. For clarity, we omit the explicit diameter dependence in $\Theta_{d_i}$ and $\sigma_{d_i}$ in this section and write $\Theta_i$ and $\sigma_{i}$ instead. The calculations are presented in terms of the full parameter set $\Theta_i$ but apply equally to the reduced parameter set  $\theta_i$. 

We use a Gaussian likelihood with multiplicative noise: the per-point standard deviation $s_{ij}(\Theta_{i})$ is proportional to the model prediction $s_{ij}(\Theta_{i}) = \sigma_{i}\,\hat{y}_{ij}$,
where $\sigma$ is the noise component of $\Theta_{i}$. This yields
\begin{equation}
P(\mathcal{D}_i \mid \Theta_{i}, \sigma_i) = \prod_{j=1}^{N_i} \mathcal{N}\!\left(y_{ij} \,\middle|\, \hat{y}_{ij},\, s_{ij}(\Theta_{i})^2\right).
\end{equation}
In the independent inference step, each diameter $d_i$ of each \textsc{emb} model is treated independently. We assign independent uniform priors to all components of $\Theta_{i}$ and sample the diameter-specific posterior,
\begin{equation}
P(\Theta_{i} \mid \mathcal{D}_i)\ \propto\ P(\mathcal{D}_i \mid \Theta_{i})\,P(\Theta_{i}).
\end{equation}
For the full model, the production prior and hyperprior ranges are summarized in Supplementary Table~\ref{tab:priors}. The reduced-model workflow uses the same likelihood, data, and phase structure, but fixed $(b_1,b_2,a_3,a_4)=(0,0,0,0)$ and infers only $(k_a,k_b,d_0,\sigma)$. These reduced independent inference runs provide the baseline single-diameter posteriors used in the main Results narrative before information is pooled hierarchically across diameters. Hierarchical pooling was performed separately within each bubble type; no pooling was performed across types. To pool information across diameters, we introduce hyperparameters for each Definity and SonoVue \textsc{emb} model
\begin{equation}
\psi = \{(\mu_k,\tau_k)\}_{k\in\mathcal{K}},\qquad
k\in\mathcal{K}=\{k_a,k_b,b_1,b_2,a_3,a_4,d_0\}
\end{equation}
with Gaussian conditional hierarchical priors for the pooled parameters 
$\Theta_{i,k}\mid\psi \sim \mathcal{N}(\mu_k,\tau_k^2)$, and uniform hyperpriors on $(\mu_k,\tau_k)_k$. For each pooled parameter $k$, the conditional hierarchical prior was truncated to the corresponding admissible range reported in Supplementary Table~\ref{tab:priors}. The observation-noise parameter $\sigma_i$ is not hierarchically pooled and is inferred independently for each diameter. \textsc{tmcmc} is then used to sample the hyperparameter posterior implied by the hierarchical model. Operationally, this phase learns the pooling priors used by the final hierarchical posteriors; it is therefore an intermediate hierarchical-learning step rather than a standalone scientific endpoint. Hierarchical pooling produces the final diameter-specific posteriors reported in this work. This yields, for each bubble $i$, a posterior conditioned on hierarchical priors inferred from pooled \textsc{emb}-model-specific data,
\begin{equation}
P(\Theta_i,\sigma_i \mid \mathcal{D}_i, \psi) \ \propto\ P(\mathcal{D}_i \mid \Theta_i,\sigma_i )\,P(\Theta_i \mid \psi)\,P(\sigma_i), \qquad \psi \sim P(\psi \mid \{\mathcal{D}_i\})
\end{equation}
where $i$ runs over the diameters of each \textsc{emb}-model. The \textsc{map} estimate was approximated by the posterior sample with the largest log-posterior value.

\textbf{\\Uncertainty Propagation}
\label{sec:meth:propagation}

\noindent Once the posterior distribution has been estimated, we propagate this uncertainty to the predicted response through the trained \textsc{dnn} surrogate. For any output quantity of interest $y$, the posterior predictive distribution is
\begin{equation}
P\!\left(y \mid \mathcal{D}_i\right)
=
\int
P\!\left(y \mid \Theta_i, \sigma_i\right)\,
P\!\left(\Theta_i, \sigma_i \mid \mathcal{D}_i\right)\,
d\Theta_i\, d\sigma_i
\;\approx\;
\frac{1}{N}\sum_{k=1}^{N}
P\!\left(y \mid \Theta_i^{(k)}, \sigma_i^{(k)}\right),
\end{equation}
where $(\Theta_i^{(k)},\sigma_i^{(k)}) \sim P(\Theta_i,\sigma_i \mid \mathcal{D}_i)$ are samples from the posterior retained for propagation. Here, $P(\Theta_i,\sigma_i \mid \mathcal{D}_i)$ denotes the posterior used for propagation, namely the final hierarchically regularized diameter-specific posterior in the hierarchical stage. In our case, $y$ is the response evaluated along the experimental grid, namely force at fixed displacement for compression and displacement at fixed force for indentation. For each posterior sample, the surrogate provides a mean response, while the observation model assigns a Gaussian uncertainty with standard deviation $\sigma_i^{(k)} \hat{y}^{(k)}$ proportional to that prediction. The posterior predictive distribution at each grid point is therefore represented as a Gaussian mixture induced by the ensemble of posterior samples. In practice, we drew $N=5{,}000$ posterior samples and evaluated the corresponding predictive distribution pointwise along the experimental reference grid. Posterior medians and central predictive intervals, including the posterior median and the pointwise 90\% posterior predictive
interval reported in this paper, were then obtained by numerical inversion of the corresponding cumulative distribution function (\textsc{cdf}) at each grid point. These posterior predictive bands are therefore fully consistent with the same heteroscedastic Gaussian likelihood used during inference, and were used to assess coverage against the experimental observations.

\newpage
\textbf{\\Continuous ranked probability score (\textsc{crps})}
\label{sec:meth:crps}

\noindent To complement interval inclusion with one proper scoring rule, we also report the \textsc{crps} for the six datasets discussed in Results~\citep{gneiting2007strictly,gneiting2007probabilistic,hersbach2000decomposition}. For a retained experimental point $j$ with observed response $y_j$ and pointwise posterior predictive \textsc{cdf} $F_j$, the score is
\begin{equation}
\mathrm{CRPS}(F_j,y_j)=\int_{-\infty}^{\infty}\left(F_j(z)-\mathbf{1}\{z\ge y_j\}\right)^2\,dz \, ,
\end{equation}
with lower values indicating better probabilistic agreement. In the present workflow, $F_j$ is the pointwise Gaussian mixture induced by the propagated posterior samples and the heteroscedastic noise model above. The posterior-predictive bands shown in the figures are obtained from pointwise quantiles of this mixture distribution, computed by numerical inversion of the corresponding \textsc{cdf}. For \textsc{crps}, we approximate the same pointwise posterior predictive distribution by Monte Carlo: for each retained posterior sample, we draw one response from its associated Gaussian component, compute the empirical ensemble \textsc{crps} at point $j$, and then average over the retained points of each curve to obtain the reported mean \textsc{crps}. The reported values are further averaged over five independent Monte Carlo replicates. This choice is attractive here because, unlike coverage alone, it rewards both calibration and sharpness under the same posterior predictive distribution.

\textbf{\\Transitional Markov Chain Monte Carlo (\textsc{tmcmc})}
\label{sec:meth:tmcmc}

\noindent Posterior sampling was performed with \textsc{tmcmc} as implemented in \textsc{korali}~\citep{martin2022korali} and executed in parallel with \textsc{mpi}. \textsc{tmcmc} builds a sequence of intermediate tempered distributions connecting prior and posterior, which improves robustness in parameter spaces that are broad, correlated, or otherwise difficult to explore directly. The numerical \textsc{tmcmc} settings used in production are reported in Supplementary Section~\ref{app:tmcmc}.

\textbf{\\Computational Implementation}
\label{sec:meth:implementation}

\noindent All workflows were executed on a high-performance computing environment consisting of multi-core \textsc{cpu} nodes and \textsc{cuda}-capable \textsc{gpu} nodes. Direct \textsc{dpd} simulations were performed with \textsc{mirheo}, while Bayesian inference used \textsc{korali} with \textsc{mpi} parallelization. Surrogate models were trained and evaluated in Python using PyTorch and standard scientific Python libraries, with all workflow settings provided through \textsc{yaml} configuration files tracked in the analysis repository.

\section*{Data and Code Availability}
The data and code used in this study are maintained in a private GitHub repository and will be made public upon publication.

\section*{Author Contributions}

B.B. developed the methodology, software development, validation, implemented the computational framework, performed the simulations and analysis, curated the data, and wrote the original draft. N.N. developed the methodology, implemented the computational framework, performed the simulations and analysis, contributed to software development, validation, curated the data, and wrote the original draft. T.P. contributed to methodology development and visualization. I.P. supervised the research. M.P. conceived the project, supervised the research, and acquired funding.

\section*{Acknowledgments}

The authors acknowledge financial support under the ERC Advanced Grant
MULTraSonicA (Grant No.\,885155) from the European Research Council. We
also acknowledge the EuroHPC Joint Undertaking for awarding the project
ID EHPC-REG-2025R02-164 access to Vega at IZUM (Slovenia).

\section*{Competing Interests}
The authors declare no competing interests.

\bibliography{refs.bib}

\clearpage
\section*{Supplementary Information}

\setcounter{section}{0}
\setcounter{subsection}{0}
\setcounter{subsubsection}{0}

\makeatletter
\renewcommand{\thesection}{S\arabic{section}}
\renewcommand{\thesubsection}{S\arabic{section}.\arabic{subsection}}
\renewcommand{\thesubsubsection}{S\arabic{section}.\arabic{subsection}.\arabic{subsubsection}}

\renewcommand{\p@subsection}{}
\renewcommand{\p@subsubsection}{}

\setcounter{figure}{0}
\renewcommand{\thefigure}{S\arabic{figure}}

\setcounter{table}{0}
\renewcommand{\thetable}{S\arabic{table}}

\setcounter{equation}{0}
\renewcommand{\theequation}{S\arabic{equation}}
\makeatother

\section{\texorpdfstring{\textsc{dpd}}{DPD} Model Implementation Details}
\label{app:dpd_details}

Each microbubble shell is represented as a triangulated surface whose vertex positions, connectivity, and reference geometric quantities are generated before the \textsc{dpd} run starts \cite{ntarakas2025dissipative}. Relevant setup scripts build a diameter-dependent shell geometry, write the corresponding parameter files, and prepare the simulation environment used later by the inference and propagation workflows. In that sense, the shell geometry used in the paper is not an abstract analytical construction separate from the codebase: it is the same geometry that is generated and consumed by the simulations. Microbubble meshes were loaded from pre-generated triangulations and immersed in a solvent domain containing explicit water and gas particles. This choice of experimental conditions aligns with force spectroscopy studies, which are conducted in liquid environments to preserve structural integrity and avoid mechanical artifacts associated with drying \cite{buchner2012}. Compression and indentation use the same shell force field but different loading prescriptions. In compression, the shell is loaded through a plate-based contact geometry and the observable of interest is the reaction force as a function of imposed displacement. In indentation, the shell is loaded by prescribed pole forces and the observable is the resulting displacement as a function of applied force. This difference is carried consistently through the entire workflow: the compression surrogates are trained in the forward direction `force(displacement, parameters)`, whereas the indentation surrogates are trained in the inverse direction `displacement(force, parameters)`. In both compression and indentation setups, simulations consisted of an equilibration phase followed by a sampling phase. Forces were averaged over the final portion of each run to reduce thermal noise. Typical simulations used time steps of $\Delta t = 0.0001$, total durations of $10^5 - 10^6$ steps, and required approximately 30--60 minutes per run on a single \textsc{nvidia} A40 \textsc{gpu}. \textsc{map} responses were evaluated at 15 displacement points per diameter. Each point used $n_\mathrm{steps} = 5{,}000$ integration steps after $n_\mathrm{eq} = 10{,}000$ equilibration steps, matching the conditions used during surrogate training data generation.

\section{Experimental force spectroscopy data}

In this work, we are using experimental force spectroscopy data from published literature. In these experiments, \textsc{afm} was used to investigate the mechanical properties of individual SonoVue\textsuperscript{\textregistered} microbubbles via indentation with a tipped cantilever. The experiments followed protocols described in detail by Morris~\cite{morris2014mechanical}. Individual MBs were imaged and targeted using an inverted optical microscope. A sharp-tipped cantilever was employed, with a nominal tip radius of 20 nm and calibrated spring constants obtained using inverse optical lever sensitivity measurements followed by the thermal noise method. The experiments were performed in aqueous conditions, using an Asylum Research MFP-1D \textsc{afm} system with Bruker MLCT AUNM tipped cantilevers (Bruker \textsc{afm} probes, Camarillo, CA) with nominal spring constants $k_c = 0.005$--$0.025$~N/m. The cantilever approach and retraction were performed at a constant speed of $3~\mu$m/s. To ensure precise measurement of local shell deformation, force spectroscopy was performed by advancing the \textsc{afm} tip toward the apex of each microbubble at a controlled rate, generating force--displacement ($F$--$\Delta$) curves. Both approach and retract curves were collected, but only the approach segment was used for fitting, as it corresponds to the compressive phase. The deformation was limited to ensure an elastic response and avoid structural damage. Since the applied force in \textsc{afm} is determined from cantilever deflection using Hooke's law, the cantilever spring constant serves only as a measurement parameter and does not influence the intrinsic membrane response. Consistent with this, experimental studies report only a weak dependence of the inferred mechanical properties on cantilever stiffness or probe geometry \cite{morris2014mechanical}. Accordingly, our simulations, which model the intrinsic force--deformation behavior of the membrane, are expected to be insensitive to the specific cantilever stiffness used in the experiments.

Tipless cantilevers were used to apply parallel-plate compression to individual Definity\textsuperscript{\textregistered} microbubbles, as described by Buchner-Santos et al.~\cite{buchner2012}. The \textsc{emb}s were immobilized at the base of poly-L-lysine coated Petri dishes via a passive float-and-adhere protocol to maintain their structural integrity. All force measurements were performed in deionized water using an Asylum Research MFP-1D \textsc{afm} system mounted on an inverted optical microscope (Nikon TE2000U). Tipless rectangular cantilevers (spring constants $k_c = 0.07$--$0.25$~N/m) with aluminum backside coating were used to compress \textsc{emb}s at a constant speed of approximately $6~\mu$m/s. The cantilever deflection and piezoelectric displacement were recorded, allowing calculation of the applied force and \textsc{emb} deformation. 

Here, we reanalyze published experimental force--displacement curves after conversion to \textsc{dpd} units and preprocessing specific to each \textsc{emb} type. Each posterior is conditioned on one representative published curve per diameter. Because the workflow repository stores only the representative curve actually used for inference, rather than the full original experimental cohorts, Table~\ref{tab:data_summary} reports the number of curves used by the inference workflow and the retained point counts, not the total number of bubbles measured in the source studies.

\begin{figure}[H]
\centering
\includegraphics[width=\textwidth]{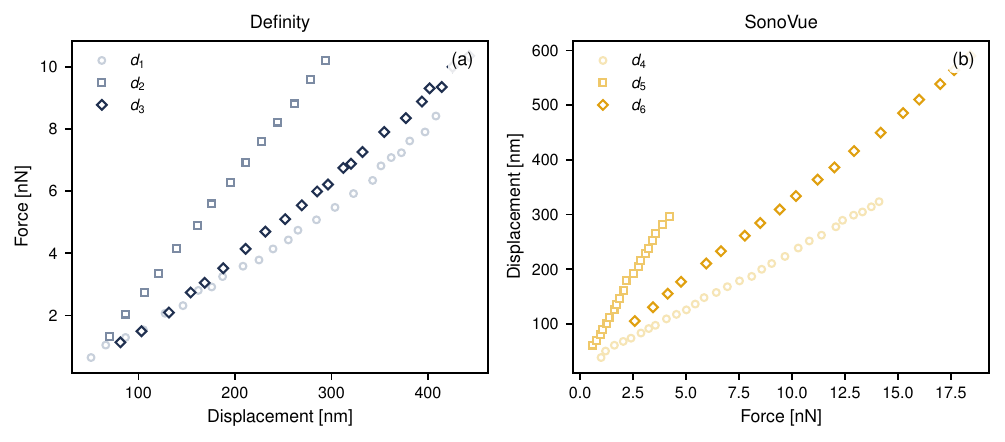}
\caption{\textbf{Experimental reference curves used throughout the calibration workflow, shown in real units.} Panel (a) shows compression data (ink) and panel (b) shows indentation data (gold). Each curve corresponds to a reprocessed published loading branch (Refs.~\cite{buchner2012,morris2014mechanical}) and serves as the diameter-specific target for surrogate evaluation, Bayesian inference, and posterior predictive propagation.}
\label{fig:reference_curves}
\end{figure}

\section{Forward-model details for hierarchical inference}
\label{app:hierarchical_derivation}

This section specifies the experiment-dependent forward maps used in the Bayesian calibration workflow. The general likelihood, prior structure, and staged hierarchical inference procedure are given in the Methods section.    

Before inference, each published reference curve is reduced to the subset that actually enters the likelihood (See Fig.~\ref{fig:reference_curves}), and Table~\ref{tab:data_summary} explicitly reports the corresponding retained point counts. In both cases of \textsc{emb}s, the retained points correspond to the linear loading regions defined in the respective reference experimental papers. The published curves are read or converted into \textsc{dpd} units and written as diameter-specific files used by the workflow, with no interpolation or additional manual alignment shift introduced afterward.

The displacement offset $d_0$ is added as an additional degree of freedom of the model, shifting all displacements so that the model encompasses the experimental uncertainty regarding the establishment of the bubble-cantilever contact. It is taken into account in an experiment-dependent way. For compression, we shift the experimental displacement before surrogate evaluation,
\begin{equation}
\tilde{x}_{ij} = \max(0, x_{ij} - d_{0,i}), \qquad
\hat{y}_{ij} = \mathcal{S}_{d_i}(\Theta_i;\tilde{x}_{ij}),
\end{equation}
where $(\mathcal{S}_{d_i})_{i=1,2,3}$ maps $(k_a,k_b,b_1,b_2,a_3,a_4,\tilde{x}) \mapsto F$.
For indentation, the surrogate predicts displacement as a function of force and parameters, and we apply the offset as an additive shift on the predicted displacement,
\begin{equation}
\hat{y}_{ij} = \hat{d}_{ij} = \mathcal{S}_{d_i}(\Theta_i;x_{ij}) + d_{0,i},
\end{equation}
where $(\mathcal{S}_{d_i})_{i=4,5,6}$ maps $(k_a,k_b,b_1,b_2,a_3,a_4,F) \mapsto d$. This asymmetry reflects the native surrogate parameterizations used in the workflow: compression surrogates predict force at a prescribed displacement, whereas indentation surrogates predict displacement at a prescribed force. In both cases, non-negativity is enforced at evaluation time by clamping predictions to zero when necessary (no negative compressive force and no negative indentation depth).

At the hierarchical stage, the bubble type hyperparameter posterior used to define the regularizing priors in the final diameter-specific inference is obtained by marginalizing over the diameter-specific parameters and curve-specific noise amplitudes:
\begin{equation}
P(\psi \mid \{\mathcal{D}_i\})
\propto
P(\psi)\prod_{i=1}^{N_d} P(\mathcal{D}_i \mid \psi),
\end{equation}
with
\begin{equation}
P(\mathcal{D}_i \mid \psi)
=
\int
P(\mathcal{D}_i \mid \Theta_i,\sigma_i)\,
P(\Theta_i \mid \psi)\,
P(\sigma_i)\,
d\Theta_i\, d\sigma_i.
\end{equation}

\subsection{\textsc{tmcmc}}
\label{app:tmcmc}
We perform Bayesian parameter inference using \textsc{tmcmc}~\citep{ching2007transitional}, a population-based algorithm that bridges the prior to the posterior through a sequence of intermediate densities,
\begin{equation}
\rho_j(\Theta_i) \propto p(\theta)^{1 - p_j}\, p(\mathcal{D}|\theta)^{p_j}, \qquad 0 = p_0 < p_1 < \cdots < p_J = 1
\end{equation}
where $p(\theta)$ denotes the prior, $p(\mathcal{D}|\theta)$ the likelihood, and $\{p_j\}$ the annealing schedule. Starting with samples drawn from the prior ($p_0=0$), \textsc{tmcmc} iteratively increases the annealing exponent to shift probability mass toward regions of high likelihood. At each stage, samples from $\rho_j$ are reweighted according to the incremental tempering step (effectively using weights proportional to $p(\mathcal{D}|\theta)^{p_{j+1}-p_j}$), resampled to control weight degeneracy, and then ``rejuvenated'' with a Markov chain kernel targeting $\rho_{j+1}$. Repeating this procedure until $p_J=1$ yields an empirical approximation of the posterior distribution.

In our implementation, \textsc{tmcmc} is provided by the \textsc{korali} framework~\citep{martin2022korali} and executed in parallel using \textsc{mpi}. \textsc{korali} manages the annealing schedule, importance reweighting and resampling between stages, and distributed evaluation of the forward model across the population. The annealing increments are selected adaptively using a target coefficient of variation (CoV) for the incremental importance weights: a larger target CoV allows larger increases in $p_j$ (fewer tempering stages but more variable weights), while a smaller target CoV enforces smaller increments (more stages and a more gradual transition). Within each stage, the Markov chain proposal is based on the (weighted) population covariance, scaled by a user-defined covariance scaling factor. This parameter controls the typical proposal step size, balancing mixing (too small leads to slow exploration) against acceptance (too large leads to frequent rejections).

The \textsc{tmcmc} settings used in this work are a population size of 50{,}000, a CoV of 0.8 (single-level inference) and 0.6 (hyperparameter inference, hierarchical-specific posteriors), and a covariance scaling parameter of 0.04 (all inference phases).

\section{Surrogates}
\label{app:surrogate}
Training data is generated by sampling the force field parameter space using Latin Hypercube Sampling \cite{joseph2020designing} and evaluating the corresponding \textsc{dpd} simulations at a prescribed set of loading points. Sampling is managed through the \textsc{korali} engine (distributed \textsc{mpi} execution) and produces diameter-specific datasets for training the surrogates. Across the six diameter-specific surrogate datasets considered here, the training sets contain approximately $5\times10^3$ to $10^4$ simulated samples per diameter. For compression, each training sample consists of a parameter vector and a sequence of displacement points with the associated simulated forces, while for indentation, each training sample consists of a parameter vector and a sequence of forces with the associated recorded displacements. In both cases, the raw simulated curves are filtered before training to remove invalid or non-physical entries to enforce consistency in the force--displacement relationship: simulated curves exhibiting abrupt force drops indicative of bubble rupture events are removed using a monotonicity-based criterion prior to architecture selection.

\begin{figure}[H]
\centering
\includegraphics[width=\textwidth]{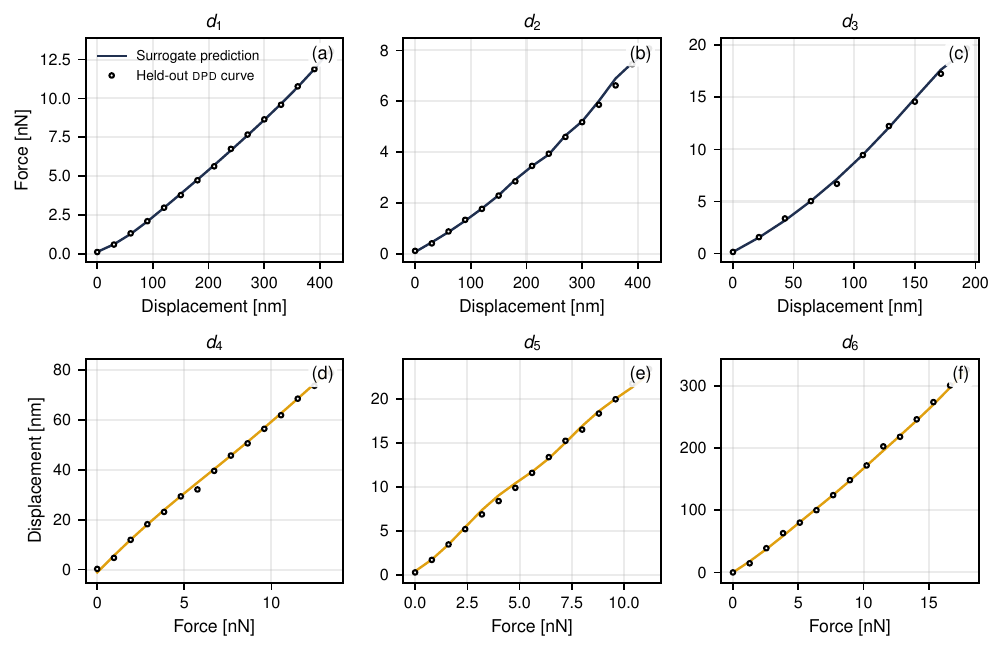}
\caption{\textbf{Grouped parameter-set holdout validation for all six surrogate models in real units.} Each panel shows one representative held-out curve chosen near the median grouped-holdout relative $L^2$ error for that diameter. Points denote held-out \textsc{dpd} responses excluded from training, and lines denote the surrogate predictions. Compression panels report reaction force as a function of displacement; indentation panels report displacement as a function of applied force. Panels (a–c) depict Definity (ink), and panels (d–f) depict SonoVue (gold).}
\label{fig:surrogate_group_holdout_examples}
\end{figure}

Each surrogate is a fully connected multilayer perceptron (\textsc{mlp}) with $\tanh$ activations and Xavier initialization \cite{glorot2010understanding} for the linear-layer weights. For each \textsc{emb} diameter, we perform an architecture sweep over a fixed set of widths and depths and select the best model based on validation performance. Specifically, we evaluate 12 candidate \textsc{mlp} spanning widths $\{32,64,128,256\}$ and depths $\{2,3,4\}$, train all candidates and retain the model with the lowest grouped-holdout median relative $L^2$ error. The selected grouped-holdout architectures are $(128,4)$, $(128,4)$, and $(128,3)$ for compression diameters 2.1, 2.9, and $3.0~\mu$m, and $(32,3)$, $(256,3)$, and $(128,4)$ for indentation diameters 3.2, 3.4, and $5.8~\mu$m, respectively. Fig.~\ref{fig:surrogate_group_holdout_examples} displays one representative held-out curve per diameter, chosen near the median grouped-holdout relative error. The points denote held-out \textsc{dpd} data excluded from training, and the lines denote the surrogate predictions. The overall conclusion is that the surrogate response is accurate on unseen parameter sets for both compression and indentation, even under the stricter curve-wise validation protocol. Inputs and targets are standardized using the feature-wise mean and standard deviation computed from the training dataset, and the normalization parameters are stored alongside the trained PyTorch \cite{ansel2024pytorch} model in a pickle file. A 90/10 split is used for training and validation. The models are trained with the Adam optimizer \cite{kingma2014adam} and \textsc{mse} loss. Training includes a validation-based learning rate reduction mechanism: when the validation loss does not improve for a fixed patience window, the learning rate is reduced by a factor of $10$ and training continues. The procedure stops after a fixed number of reduction rounds or when the maximum number of epochs is reached. Unless otherwise stated, training is performed on \textsc{cpu}s. For reduced-model analyses in which $(b_1,b_2,a_3,a_4)$ are fixed to zero, we reuse the same trained surrogates and evaluate them with these inputs set to zero, rather than retraining a separate reduced surrogate.

\newpage
\section{Sensitivity analysis}
\label{app:sensitivity}
First-order Sobol indices were computed using \textsc{SALib}
\cite{herman2017salib} via the Saltelli estimator with $N = 2,048$
samples drawn over the prior ranges listed in Table~\ref{tab:priors}, using the full surrogate (with nonlinear terms
active). Indices were computed pointwise along the loading branch and
then averaged over the experimentally explored range, as described in
Section~\ref{sec:res:sensitivity}.

\subsection{Parameters and validation tables}
\begin{table}[H]
\centering
\caption{\textbf{Surrogate validation summary for all six diameters.} We report only the grouped-holdout median relative $L^2$ error from the stricter parameter-set holdout audit, in which whole simulated curves are excluded from training.}
\label{tab:surrogate_validation}
{
\footnotesize
\setlength{\tabcolsep}{5pt}
\begin{tabular}{@{}lcc@{}}
\toprule
\textbf{Setup} & \textbf{Diameter [$\mu$m]} & \textbf{Grouped-holdout median rel. $L^2$ [\%]} \\
\midrule
 & 2.1 & 0.92 \\
Compression & 2.9 & 2.09 \\
 & 3.0 & 2.06 \\
 \hline
 & 3.2 & 2.20 \\
Indentation & 3.4 & 2.55 \\
 & 5.8 & 1.55 \\
\bottomrule
\end{tabular}
}
\end{table}

\begin{table}[H]
\centering
\caption{\textbf{Experimental datasets used for inference.} Each posterior is conditioned on one reprocessed published curve per diameter. ``Source points'' denotes the number of points parsed from the published/source curve before \textsc{dpd}-unit conversion and trimming. ``Retained points'' denotes the number of points in the processed \texttt{.dat} curve actually used by the inference workflow. Source references are cited in the table; dataset DOIs are provided where available.}
\label{tab:data_summary}
\resizebox{\textwidth}{!}{%
\begin{tabular}{lccccccc}
\toprule
\textbf{Type} & \textbf{Diameter [$\mu$m]} & \textbf{Source reference} & \textbf{Commercial agent} & \textbf{Curves used} & \textbf{Source points} & \textbf{Retained points} & \textbf{Input type} \\
\midrule
 & 2.1 &  &  & 1 & 27 & 24 &  \\
Compression & 2.9 & Buchner-Santos et al.~\cite{buchner2012} & Definity & 1 & 17 & 14 & Reprocessed \\
 & 3.0 &  &  & 1 & 25 & 22 & \\
 \hline
 & 3.2 &  &  & 1 & 32 & 29 &  \\
Indentation & 3.4 & Morris~\cite{morris2014mechanical} & SonoVue & 1 & 46 & 20 & Reprocessed \\
 & 5.8 &  &  & 1 & 22 & 19 &  \\
\bottomrule
\end{tabular}
}
\end{table}

\begin{flushleft}
\footnotesize
Dataset DOIs: Buchner-Santos et al. (2012): 10.1021/la204801u.
\end{flushleft}

\begin{table}[H]    
\centering
\caption{\textbf{Full-model production prior and hyperprior ranges used in the compression and indentation hierarchical inference workflows.} The reduced model uses the same workflow and reference data, but fixes $(b_1,b_2,a_3,a_4)=(0,0,0,0)$ and infers only $(k_a,k_b,d_0,\sigma)$.}
\label{tab:priors}
\resizebox{\textwidth}{!}{%
\begin{tabular}{lcccccc}
\toprule
\textbf{Parameter} &
\textbf{Prior (compression)} &
\textbf{Hyperprior $\boldsymbol{\mu}$ (compression)} &
\textbf{Hyperprior $\boldsymbol{\sigma}$ (compression)} &
\textbf{Prior (indentation)} &
\textbf{Hyperprior $\boldsymbol{\mu}$ (indentation)} &
\textbf{Hyperprior $\boldsymbol{\sigma}$ (indentation)} \\
\midrule
$k_a$ & [$5.58{\times}10^3$, $2.79{\times}10^4$] & [$5.58{\times}10^3$, $2.79{\times}10^4$] & [$0$, $2.79{\times}10^4$] & [$5.58{\times}10^2$, $5.58{\times}10^5$] & [$5.58{\times}10^2$, $5.58{\times}10^5$] & [$0$, $2.79{\times}10^5$] \\
$k_b$ & [$100$, $1{,}000$] & [$100$, $1{,}000$] & [$0$, $500$] & [$400$, $70{,}000$] & [$400$, $70{,}000$] & [$0$, $35{,}000$] \\
$b_1$ & [$0$, $3$] & [$0$, $3$] & [$0$, $3$] & [$0$, $3$] & [$0$, $3$] & [$0$, $3$] \\
$b_2$ & [$0$, $10$] & [$0$, $10$] & [$0$, $10$] & [$0$, $10$] & [$0$, $10$] & [$0$, $10$] \\
$a_3$ & [$-2.5$, $3$] & [$-2.5$, $3$] & [$0$, $5$] & [$-2.5$, $3$] & [$-2.5$, $3$] & [$0$, $5$] \\
$a_4$ & [$0$, $4$] & [$0$, $4$] & [$0$, $4$] & [$0$, $4$] & [$0$, $4$] & [$0$, $4$] \\
$d_0$ [\textsc{dpd}] & [$0$, $0.5$] & [$0$, $0.5$] & [$0$, $0.3$] & [$-0.5$, $0.5$] & [$-0.5$, $0.5$] & [$0$, $0.3$] \\
$\sigma$ & [$0$, $1$] & --- & --- & [$0$, $1$] & --- & --- \\
\bottomrule
\end{tabular}
}
\end{table}

\begin{table}[H]
\centering
\caption{\textbf{Curve-averaged posterior-predictive \textsc{crps} for the reduced and full workflows across the six datasets reported in the paper.} Lower values indicate better probabilistic agreement. Compression \textsc{crps} is reported in force units (nN), and indentation \textsc{crps} is reported in displacement units (nm). Values are averaged over five independent Monte Carlo replicates.}
\label{tab:crps_trusted}
{
\footnotesize
\setlength{\tabcolsep}{5pt}
\begin{tabular}{@{}lcccc@{}}
\toprule
\textbf{Type} & \textbf{Diameter [$\mu$m]} & \textbf{Reduced mean \textsc{crps}} & \textbf{Full mean \textsc{crps}} & \textbf{Units} \\
\midrule
 & 2.1 & 0.087 & 0.080 & nN \\
Compression & 2.9 & 0.071 & 0.060 & nN \\
 & 3.0 & 0.064 & 0.055 & nN \\
 \hline
 & 3.2 & 1.817 & 1.919 & nm \\
Indentation & 3.4 & 3.181 & 2.411 & nm \\
 & 5.8 & 2.967 & 3.054 & nm \\
\bottomrule
\end{tabular}
}
\end{table}

\begin{table}[H]
\centering
\caption{\textbf{Per-diameter 90\% posterior predictive coverage for the
reduced hierarchical model.} Each value is the fraction of experimental
points that fall within the 90\% pointwise posterior predictive band.}
\label{tab:coverage}
\begin{tabular}{lcccc}
\toprule
\textbf{Setup} & \textbf{Diameter [$\mu$m]} & \textbf{Coverage (Full) [\%]} & \textbf{Coverage (Reduced) [\%]} \\
\midrule
 & 2.1 & 95.8 & 95.8 \\
Compression & 2.9 & 100.0 & 100.0 \\
 & 3.0 & 90.9 & 100.0 \\
 \hline
 & 3.2 & 96.6 & 93.1 \\
Indentation & 3.4 & 95.0 & 95.0  \\
 & 5.8 & 94.7 & 94.7 \\
\bottomrule
\end{tabular}
\end{table}

\begin{table}[H]
\centering
\caption{\textbf{Final \textsc{map} parameter values for all models.} All elastic parameters ($k_a$, $k_b$, $\mu$) taken from Eq.~\ref{stretch_ene} are in \textsc{dpd} units; $d_0$ is in \textsc{dpd} length units. }
{
\footnotesize
\setlength{\tabcolsep}{5pt}
\begin{tabular}{@{}lllllllllll@{}}
\toprule
Commercial agent & Model & Diameter & $k_a$ & $k_b$ & $\mu$ & $b_1$ & $b_2$ & $a_3$ & $a_4$ & $d_0$ \\
\midrule
 & Full & $d_{1}$ & 7,481 & 196.7 & 2,494 & 1.812 & 2.013 & -1.422 & 0.9323 & 0.0008722 \\
 & Reduced & $d_{1}$ & 7,843 & 192.1 & 2,614 & -- & -- & -- & -- & $2.573 \times 10^{-5}$ \\
Definity & Full & $d_{2}$ & $1.789 \times 10^{4}$ & 959.1 & 5,963 & 0.8105 & 0.0667 & -1.996 & 3.468 & 0.1212 \\
 & Reduced & $d_{2}$ & $1.88 \times 10^{4}$ & 915.8 & 6,266 & -- & -- & -- & -- & 0.109 \\
 & Full & $d_{3}$ & $1.197 \times 10^{4}$ & 489.7 & 3,991 & 1.715 & 5.425 & -1.223 & 3.585 & 0.1132 \\
 & Reduced & $d_{3}$ & $1.38 \times 10^{4}$ & 432.5 & 4,599 & -- & -- & -- & -- & 0.1139 \\
 \hline
 & Full & $d_{4}$ & $1.26 \times 10^{4}$ & $1.431 \times 10^{4}$ & 4,199 & 2.349 & 2.512 & -0.5179 & 1.954 & 0.09217 \\
 & Reduced & $d_{4}$ & $1.501 \times 10^{4}$ & $1.257 \times 10^{4}$ & 5,003 & -- & -- & -- & -- & 0.1021 \\
SonoVue & Full & $d_{5}$ & 6,065 & 4,314 & 2,022 & 1.822 & 6.299 & 2.423 & 1.282 & 0.1112 \\
 & Reduced & $d_{5}$ & 1,656 & $1.274 \times 10^{4}$ & 552 & -- & -- & -- & -- & 0.08105 \\
 & Full & $d_{6}$ & 9,440 & $3.091 \times 10^{4}$ & 3,147 & 1.596 & 1.417 & -1.286 & 2.995 & 0.2448 \\
 & Reduced & $d_{6}$ & $2.077 \times 10^{4}$ & $1.264 \times 10^{4}$ & 6,922 & -- & -- & -- & -- & 0.2325 \\
\bottomrule
\end{tabular}
}
\end{table}

\end{document}